\documentclass[lettersize,journal]{IEEEtran}
\usepackage{amsfonts,color}
\usepackage{algorithm}
\usepackage{array}
\usepackage[caption=false,font=normalsize,labelfont=sf,textfont=sf]{subfig}
\usepackage{textcomp}
\usepackage{stfloats}
\usepackage{url}
\usepackage{verbatim}
\usepackage{graphicx}
\usepackage{cite}
\newtheorem{remark}{Remark}
\usepackage[cmex10]{amsmath}

\usepackage{bbm}
\usepackage{upgreek}
\usepackage[normalem]{ulem}
\usepackage{mathtools}

\usepackage{graphicx}
\usepackage{amssymb}
\newcommand*{\rom}[1]{\expandafter\romannumeral #1}
\usepackage{epsfig}
\usepackage{epstopdf}
\usepackage{color}
\usepackage{algpseudocode}
\makeatletter
\def\BState{\State\hskip-\ALG@thistlm}
\makeatother
\usepackage{multirow}
\usepackage{hyperref}
\usepackage{enumitem}
\usepackage[utf8]{inputenc}
\usepackage{nomencl}

\allowdisplaybreaks
\makenomenclature
\usepackage{etoolbox}
\renewcommand\nomgroup[1]{%
  \item[\bfseries
  \ifstrequal{#1}{A}{Index sets}{%
  \ifstrequal{#1}{D}{Parameters (partial list)}{%
  \ifstrequal{#1}{H}{Variables for vehicles}{%
  \ifstrequal{#1}{L}{Variables for DERs}{%
  \ifstrequal{#1}{Q}{Variables for hydrogen fueling station}{%
  \ifstrequal{#1}{T}{Constants and Parameters}{} } } } } }%
]}
\usepackage{framed} 
\usepackage{multicol} 
\renewcommand*\nompreamble{\begin{multicols}{2}}
\renewcommand*\nompostamble{\end{multicols}}
\usepackage{array}
\usepackage{makecell} 
\usepackage{tabularx}
\usepackage{xcolor}
\usepackage{lipsum}
\usepackage{ bbold }

\begin{document}

\title{Integer-Clustering Optimization of Hydrogen and Battery EV Fleets Considering DERs}
        
\author{Sijia Geng$^*$,~\IEEEmembership{Member,~IEEE,} Thomas Lee$^*$,~\IEEEmembership{Student Member,~IEEE,}
Dharik Mallapragada,
and Audun Botterud,~\IEEEmembership{Member,~IEEE}
\thanks{This work was supported by the Ralph O’Connor Sustainable Energy Institute (ROSEI) at Johns Hopkins University and MIT Energy Initiative. (Corresponding author: Sijia Geng, \texttt{sgeng@jhu.edu}).}
\thanks{Sijia Geng is with the Department of Electrical and Computer Engineering,
Johns Hopkins University, Baltimore, MD, 21218, USA. Thomas Lee is with MIT Institute for Data, Systems, and Society, Massachusetts Institute
of Technology, Cambridge, MA 02139, USA. Dharik Mallapragada is with Chemical and Biomolecular Engineering Department, Tandon School of Engineering, New York University, Brooklyn, NY 11201, USA. Audun Botterud is with the Laboratory for Information and Decision
Systems, Massachusetts Institute of Technology, Cambridge, MA 02139 USA. }
\thanks{$^*$The first two authors contribute equally.}
}



\maketitle

\nomenclature[A,01]{$s\in[S]$}{Representative days.}
\nomenclature[A,02]{$t\in[T_d]$}{Time intervals within each representative day.}
\nomenclature[A,03]{$k\in[K]$}{Trip blocks. $[K]=\cup_{s\in[S]} \mathcal{K}_s$, where $\mathcal{K}_s$ correspond to each representative day.}
\nomenclature[A,04]{$i\in[I]$}{Vehicle types.}
\nomenclature[A,05]{$j\in[J]$}{Charger types.}
\nomenclature[A,06]{$l\in[L]$}{Season groups for demand charge. $[S]=\cup_{l\in[L]} \mathcal{L}_l$, where $\mathcal{L}_l$ corresponds to groups of representative days with the same demand charge.}
\nomenclature[A,07]{$\mathcal{T}_\text{del}$}{Hours of the day when hydrogen delivery can be scheduled.}

\nomenclature[D,00]{$N_s^{\text{d}}$}{Weighting (days per year) assigned to representative day $s$.}
\nomenclature[D,01]{$D_k$}{Travel distance (km) of trip block $k$.}
\nomenclature[D,02]{$\eta^{\text{v}}_{ki}$}{Driving efficiency (kWh/km) for block $k$, if using type-$i$ vehicle.}
\nomenclature[D,05]{$A_{k}(t)$}{Indicator of en-route (1) or idle (0) status of trip block $k$.}
\nomenclature[D,06]{$U_{k}(t)$}{Indicator of trip block $k$'s starting time interval.}
\nomenclature[D,07]{$V_{k}(t)$}{Indicator of trip block $k$'s return time interval.}
\nomenclature[D,08]{$P_{ij}$}{Charging capacity (kW) per vehicle-charger type combination.}
\nomenclature[D,09]{$R_i$}{Vehicle range expressed as energy capacity (kWh).}
\nomenclature[D,11]{${a}^\text{pv}_s(t)$}{Intermittent capacity factor for solar PV.}

\nomenclature[H,01]{$N_i^\text{v}$}{Number of type-$i$ vehicles purchased.}
\nomenclature[H,02]{$N_j^\text{c}$}{Number of type-$j$ chargers installed.} 
\nomenclature[H,03]{$b_{ki}$}{Assignment indicator (0,1) of block $k$ to vehicle type-$i$.} 
\nomenclature[H,04]{$d_{ki}$}{Onboard stored energy (kWh) when block $k$ departs.} 
\nomenclature[H,05]{$n_{is}(t)$}{Number of type-$i$ vehicles at the depot.}
\nomenclature[H,06]{$m_{ijs}(t)$}{Number of type-$i$ vehicles undergoing charging with type-$j$ chargers.}
\nomenclature[H,07]{$p^{\text{v}}_{is}(t)$}{Depot-pooled electric charging power (kW) to type-$i$ vehicles.}
\nomenclature[H,08]{${p}^{\text{h}}_s(t)$}{Depot-pooled hydrogen charging power (kW).}
\nomenclature[H,09]{${{p}}_s^\text{df}(t)$}{Depot-pooled diesel fuel charging power (kW).}
\nomenclature[H,10]{$q_{is}(t)$}{Depot-pooled state-of-energy (kWh) per vehicle type.}
\nomenclature[H,11]{$p^\text{pk}_{l}$}{Peak depot-pooled electric power (kW) in season group $l$.}

\nomenclature[L,00]{$\overline{p}^\text{pv}$}{Power capacity (kW) of solar PV generation.}
\nomenclature[L,02]{$\overline{p}^\text{b}$}{Power capacity (kW) of battery.}
\nomenclature[L,03]{$\overline{e}^\text{b}$}{Energy capacity (kWh) of battery.}
\nomenclature[L,04]{${p}^\text{g}_s(t)$}{Electric power (kW) supplied by the grid.}
\nomenclature[L,05]{${p}^\text{d}_s(t)$}{Depot-pooled electric power demand (kW) of the BEVs.}
\nomenclature[L,06]{${p}^\text{curt}_s(t)$}{Curtailment (kW) from renewable resources.}
\nomenclature[L,07]{${p}^{\text{b}+}_s(t)$}{The grid withdrawal power (kW) for battery charging.}
\nomenclature[L,08]{${p}^{\text{b}-}_s(t)$}{The grid injection power (kW) from battery discharging.}
\nomenclature[L,09]{${e}^\text{b}_s(t)$}{State-of-energy (kWh) of battery storage.}
\nomenclature[L,10]{$\gamma_s(t)$}{Binary indicator for battery charging (1) or discharging (0).}
\nomenclature[Q,01]{$\overline{w}^{\text{h}}$}{Capacity (kg) of the (low-pressure) hydrogen tank.}
\nomenclature[Q,02]{$\overline{w}^{\text{bf}}$}{Capacity (kg) of the (high-pressure) storage buffer.}
\nomenclature[Q,03]{$\overline{p}^{\text{elz}}$}{Maximum electric power consumption (kW) of the electrolyzer.}
\nomenclature[Q,04]{$\overline{p}^{\text{lcpr}}$}{Maximum electric power consumption (kW) of low-pressure compressor.}
\nomenclature[Q,05]{$\overline{p}^{\text{cpr}}$}{Maximum electric power consumption (kW) of the compressor.}
\nomenclature[Q,06]{$\overline{p}^{\text{cl}}$}{Maximum electric power consumption (kW) of the cooling system.}
\nomenclature[Q,07]{${w}^\text{del}_s(t)$}{Mass of hydrogen (kg) delivered via truck.}
\nomenclature[Q,08]{${w}^{\text{h}}_s(t)$}{Mass of hydrogen (kg) stored in the hydrogen tank.}
\nomenclature[Q,09]{${w}^\text{bf}_s(t)$}{Mass of hydrogen (kg) stored in the high-pressure storage buffer.}
\nomenclature[Q,10]{${p}^\text{elz}_s(t)$}{Electric power (kW) consumed by electrolyzer.} 
\nomenclature[Q,11]{${p}^\text{lcpr}_s(t)$}{Electric power (kW) consumed by low-pressure compressor to transport hydrogen into the (low-pressure) hydrogen tank.}
\nomenclature[Q,12]{${p}^\text{cpr}_s(t)$}{Electric power (kW) consumed by the compressor between the low-pressure hydrogen tank to the high-pressure storage buffer.}
\nomenclature[Q,13]{${p}^\text{cl}_s(t)$}{Electric power (kW) consumed by the cooling system when dispensing hydrogen to the vehicle.}
\begin{table*}[!t]   
\begin{framed}
\printnomenclature
\end{framed}
\end{table*}

\begin{abstract}
Electrified transportation leads to a tighter integration between transportation and energy distribution systems. In this work, we develop scalable optimization models to co-design hydrogen and battery electric vehicle (EV) fleets, distributed energy resources, and fast-charging and hydrogen-fueling infrastructure to efficiently meet transportation demands. A novel integer-clustering formulation is used for optimizing fleet-level EV operation while maintaining accurate individual vehicle dispatch, which significantly improves the computation efficiency with guaranteed performance. We apply the optimization model to Boston’s public transit bus network using real geospatial data and cost parameters. Realistic insights are provided into the future evolution of coupled electricity-transportation-hydrogen systems, including the effects of electricity price structure, hydrogen fuel cost, carbon emission constraint, temperature effects on EV range, and distribution system upgrade cost.
\end{abstract}

\begin{IEEEkeywords}
Electric vehicles, hydrogen fuel cell vehicles,
distributed energy resources, optimization, smart charging, integer clustering.
\end{IEEEkeywords}

\section{Introduction}

\IEEEPARstart{E}{lectric}
vehicles (EVs) offer several advantages over conventional vehicles, including lower operating and maintenance costs, improved performance and efficiency, and reduced carbon emissions. Numerous works have studied the planning and operation of EVs and their supporting energy infrastructure \cite{muratori2018impact,wei2021personal,BURNHAM2017237,bao2023optimal}. 
While light-duty EVs in residential settings have received considerable attention, heavy-duty EV fleets are less studied. In addition, most existing research has focused on battery EVs (BEVs), driven by declining battery costs and increasing market share. However, hydrogen fuel cell EVs (FCEVs) may be a promising alternative in the future-- especially for heavy-duty applications--due to their potential advantages in range and refueling time \cite{mai2018electrification}. 
We aim to compare the BEV and FCEV technologies and quantify the impacts of heavy-duty EV charging on the electricity demand and energy infrastructure needs, considering the potential of distributed energy resources (DERs)\cite{geng2020chance}. We focus on modeling a large number of
commercial EV fleets, such as public buses, school buses, and delivery
freight, and develop scalable optimization tools for planning and operation of EVs and the energy infrastructure.  

To fully examine the complex interactions of multi-energy systems \cite{geng2020optimal}, including DERs and hydrogen infrastructure, and large-scale fleets and chargers, can be intractable \cite{bertossi1987some}, especially for the planning problem.  
Therefore, previous works typically make highly simplified restrictions that can lead to inefficient solutions. Public transit contrasts with both personal vehicles and the fleet type considered in \cite{borlaug2021heavy}, which are characterized by a relatively unimodal driving profile during the day. In comparison, transit schedules are characterized by numerous return-to-depot type trip sequences made throughout the day, resulting in a more complex feasible set of charging strategies.  
\cite{li2019mixed} studied an EV fleet scheduling problem with fleet sizing decision, though, also used a fixed charging rule and discretized energy levels. \cite{bi2018wireless} proposed a multi-objective life cycle optimization model to deploy wireless charging infrastructure for BEVs, with a focus on the siting of chargers within the bus network. In the same vein, \cite{kim2020development} studied the deployment of hydrogen fueling stations on a nationwide scale in Korea. 
\cite{pham2022techno} studied the value of DERs for meeting the energy demand of heavy-duty EVs, taking the EV charging load as a fixed input instead of decision variables. 
Operation and scheduling problems have been a focus of the vast literature.
\cite{al2023multi} proposed a multi-battery flexibility model to conservatively estimate the aggregate flexibility set of EVs using a few virtual batteries. However, the total number and composition of the individual EVs are given parameters. \cite{botkin2021distributed} presented a distributed controller for EV charging using aggregated fleet energy; however, the trip assignments were taken as given input parameters, and the vehicle composition was homogeneous. Methods that use a space-time-energy graph \cite{bunte2009overview,li2019mixed} often discretize the battery charging levels, which may reduce the accuracy in considering energy cost.

With the increasing number of EVs on the roads, there is a greater need for EV charging stations and hydrogen fueling stations.
In this paper, we focus on the design of the charging depot and EV fleets to highlight the interaction with energy distribution systems. 
To fill in the gaps of the literature, we treat investments and operations of the fleet and energy infrastructure as decision variables to be optimized, and aim to minimize the total cost and carbon emissions.
Towards this end, we formulate hydrogen infrastructure, including on-site hydrogen production and storage, and DERs \cite{geng2020optimal}, such as solar photovoltaic (PV) and battery energy storage system (BESS), at the depot. 
 One of the key challenges in such a study lies in the large problem size and complexity. The integer-clustering formulation, as first proposed by the authors in \cite{geng2024integer}, innovatively models vehicle operations (i.e., charging and fueling) at the fleet level to reduce the computational requirement compared to formulations based on individual vehicles. Only vehicle and charger types are differentiated to capture different driving and charging efficiencies, yet, guaranteed accuracy and significant speed-ups were proved.  
In this paper, we extend the method to model BEV, FCEV, and hybrid diesel fleets, and develop an optimization model to determine the optimal design (investment plan) and operation for vehicles and the corresponding energy infrastructure at the depot.
Other strategies to meet the EV demand, in particular, incentivizing demand flexibility through time-of-use (TOU) electricity prices, are studied. Moreover, a sensitivity analysis is performed to analyze the impacts of hydrogen fuel price and carbon emissions limit on the optimal design. The model also considers practical issues such as the effects of cold weather on vehicle driving range and distribution system update.

The following contributions are made: 1) The important characteristics and coupling of three key sectors, electricity, hydrogen, and transportation, are modeled in this paper. 2) We leverage an efficient integer-clustering method and formulate a scalable optimization problem to identify the optimal investment plan for EV fleets and supporting energy infrastructure while considering accurate operations. 
3) The model evaluates BEV and FCEV technologies, and the potential of DERs in minimizing the cost and carbon emissions.
4) Realistic insights are provided based on a real-world case study of Boston's public transit bus network, using a real geospatial dataset for bus schedules and actual techno-economic parameters. 
The rest of the paper is organized as follows: Problem description is provided in Section~\ref{sec:problem}. Section~\ref{sec:models_trans} presents the integer-clustering modeling for EV fleets, and Section~\ref{sec:models} models the energy infrastructures. Section~\ref{sec:results} provides case studies and analysis, and Section~\ref{sec:conclusion} concludes the paper.

\section{Problem Description}\label{sec:problem}
This section summarizes the optimization problem. The problem aims to find the least-cost investment plan for EV fleets and corresponding energy infrastructures that satisfy a set of operational constraints.

\begin{align}
&(\mathcal{P}_1) \ \ \min\limits_{\boldsymbol{X}_
\text{agg}}\ J_\text{obj} \ \ \eqref{eq:objective} \nonumber \\
& \mathrm{subject\ to} \nonumber \\
&\quad \text{Integer-clustering-based fleet constraints}\ \  \eqref{eq_block}-\eqref{eq_m_positive},\nonumber\\
&\quad \text{Electrical system constraints}\ \ \eqref{eq_power_balance_elec}-\eqref{eq_non_neg},\nonumber\\
&\quad \text{Hydrogen station constraints}\ \ \eqref{dynamic_h2}-\eqref{eq_h_posi},\nonumber\\
&\quad \text{Carbon emission constraint}\ \ \eqref{eq_carbon},\nonumber\\
&\quad \text{Coupling constraints}\ \ \eqref{eq_couple_pd}-\eqref{eq_couple_phh}.\nonumber
\end{align} 
The overall formulation is a deterministic\footnote{The trip schedules are modeled with given deterministic start and end times motivated by the use case of public transit fleets. The uncertainty in renewable generation is accounted for through a scenario approach.} mixed integer linear program (MILP).
The decision variables ${\boldsymbol{X}_
\text{agg}}$ are summarized in the nomenclature table (grouped by vehicles, DERs, and hydrogen components). 
The objective function sums all costs, including investment costs and operation costs, 
\begin{flalign}
    J_\text{obj}& =\sum_i N^{\text{v}}_i c^{\text{v}}_i
    +\sum_{j} N^{\text{c}}_j c^{\text{c}}_j + \sum_i  \sum_k D_{k} b_{ki} c^{\text{m}}_i 
\nonumber&&
\\
& \hspace{-0.6cm} + \sum_l  p^{\text{pk}}_{l} c^{\text{pk}}_{l}  + \sum_s \! \sum_{t\in [T_d]} \! N_s^{\text{d}}\Delta T \big( p_s^{\text{g}}(t) c_s^\text{g}(t) + {p}_s^\text{df}(t) c_s^\text{df}(t) \big)
 \nonumber&&
 \\
 & \hspace{-0.6cm} + \sum_s \sum_{t\in [T_d]} N_s^{\text{d}} {w}_s^\text{del}(t) c_s^\text{del}(t) +  \sum_s
 \big(w^\text{h}_s(1) + w^\text{bf}_s(1)\big) c_s^\text{del}(1)\nonumber&& \\
& \hspace{-0.6cm} +
\overline{p}^{\text{s}} c^{\text{s}}+
\overline{e}^{\text{b}} c^{\text{b}} + 
\overline{w}^{\text{h}} c^{\text{h}} + \overline{w}^{\text{bf}} c^{\text{bf}} \nonumber&&\\
& \hspace{-0.6cm} + 
\frac{ \overline{p}^{\text{elz}} }{C^{\text{elz}}} c^{\text{elz}}  + \frac{ \overline{p}^{\text{lcpr}}}{C^{\text{lcpr}}} c^{\text{lcpr}} +
\frac{ \overline{p}^{\text{cpr}} }{C^{\text{cpr}}} c^{\text{cpr}} +
\frac{ \overline{p}^{\text{cl}} }{C^{\text{cl}}} c^{\text{cl}}.&&\label{eq:objective}
\end{flalign}
The indices $i,j,k,l,s,t$ are explained in the nomenclature. The cost parameters $c$ and efficiency parameters $C$, together with their meanings, are listed in Tables~\ref{table:EV_parameter}-~\ref{table:all_tech_parameter}.
$N^{\text{v}}_i c^{\text{v}}_i$ and $N^{\text{c}}_j c^{\text{c}}_j$ are the capital cost of EVs and chargers. $D_{k} b_{ki} c^{\text{m}}_i$ is the maintenance cost of the fleet, which depends on the decision $b_{ki}$ of vehicle dispatch. $p^{\text{pk}}_{l} c^{\text{pk}}_{l}$ and $N_s^{\text{d}} \Delta T p_s^{\text{g}}(t) c_s^\text{g}(t)$ denote the (peak) demand charge and energy cost of the electricity supplied by the grid, respectively. $N_s^{\text{d}}\Delta T {p}_s^\text{df}(t)  c_s^\text{df}(t)$ captures the cost of diesel fuel for conventional vehicles. $N_s^{\text{d}}{w}_s^\text{del}(t) c_s^\text{del}(t)$ captures the cost of hydrogen fuel by delivery, and $
 \big(w^\text{h}_s(1) + w^\text{bf}_s(1)\big) c_s^\text{del}(1)$ is the cost of initial hydrogen stored in the hydrogen tank and storage buffer. The second-to-last line in \eqref{eq:objective} are the capital costs of the various components, including the solar panel, battery, hydrogen tank, and buffer. The last line denotes the capital costs of the electrolyzer, low-pressure compressor, compressor, and cooling system. 

\section{Modeling of Vehicle Fleets}\label{sec:models_trans}

\subsection{Depot-Based Scheduling Structure and Time Indexing}

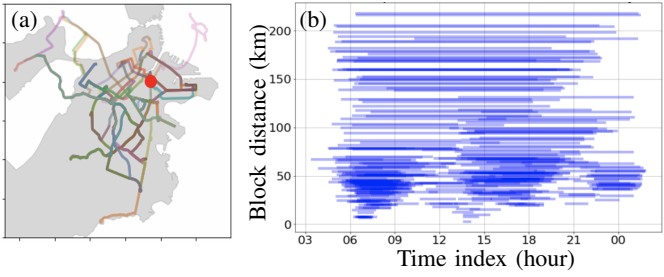
\begin{figure}[]
\begin{center}
\begin{picture}(245.0, 95.0)
\put(  -7,  0){\epsfig{file=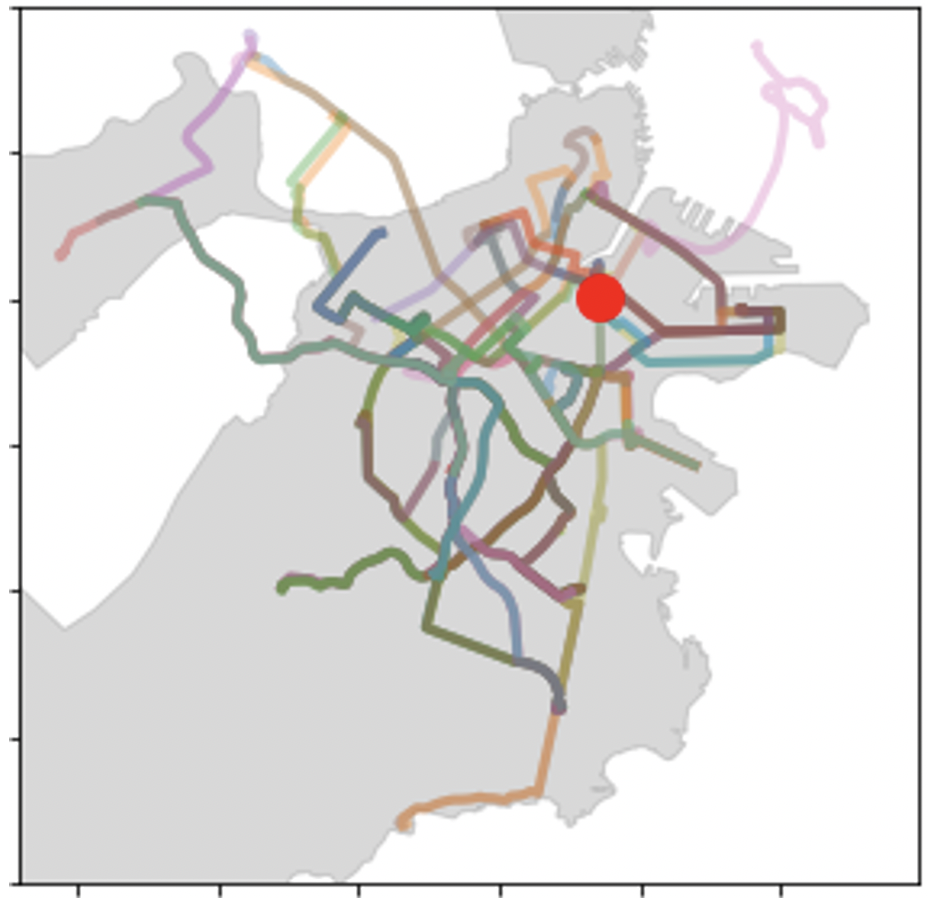,height=.176\textwidth,width=.170\textwidth}}      
\put(  94,  -2){\epsfig{file=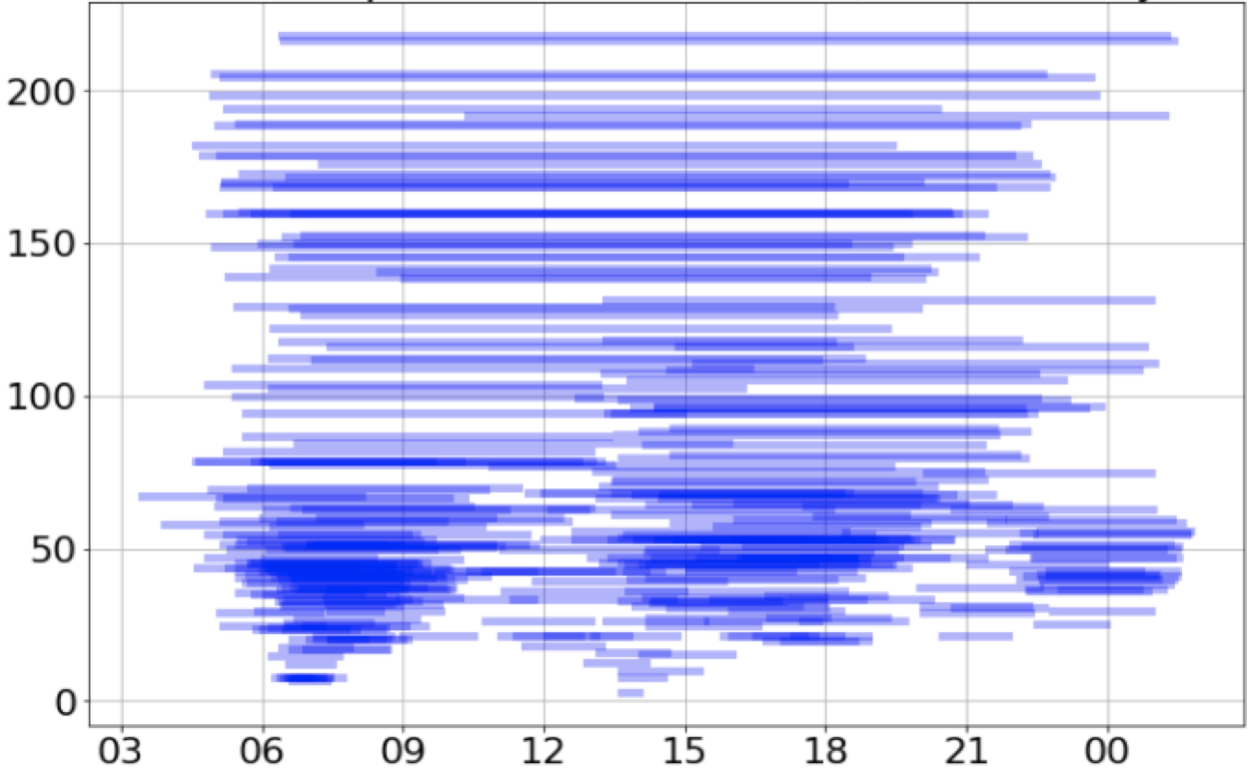,height=.180\textwidth}}  

\put( 87, 8){\small\rotatebox{90}{Block distance (km)}}
\put( 143, -9){\small{Time index (hour)}}

\put( -3, 81.5){\small{(a)}}
\put( 106, 81.5){\small{(b)}}
\end{picture}
\end{center}
      \caption{{(a) Geographical location of the Cabot bus depot located in south Boston and the bus routes supported by the depot. (b) Block schedules for Cabot depot on a representative weekday in Fall season.}}\label{fig:cabot_schedules_a}
       \vspace{-0.05cm}
\end{figure}

\begin{figure}[h]
    \centering
    \includegraphics[width=0.5\linewidth]{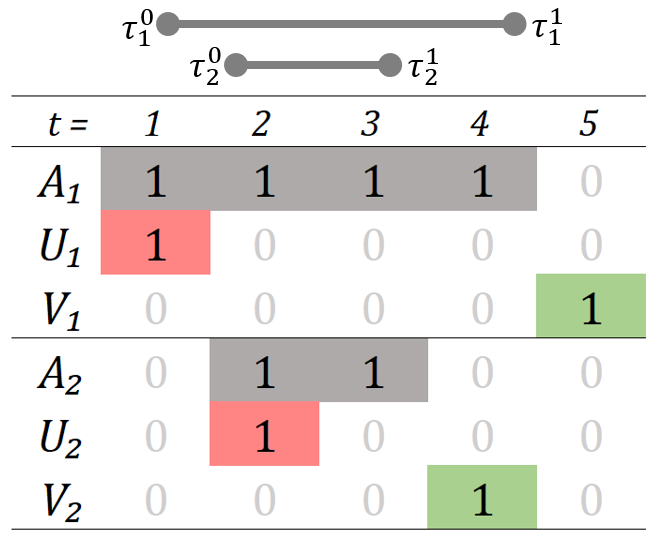}
    \caption{Illustration of discrete-time convention for two trips with continuous-time intervals $[\tau^0_1, \tau^1_1]$ and $[\tau^0_2, \tau^1_2]$, and $A,U,V$ matrices for each trip blocks.}
    \label{fig:time_convention}
\end{figure}

We focus on a single depot, as shown by the red dot in Fig.~\ref{fig:cabot_schedules_a}\,(a), where transport demand is represented by ``trip blocks", as shown by the curves. Each trip block is serviced by a single vehicle, which begins at the depot, drives through a sequence of trip segments, and ends at the depot \cite{wessel2017constructing}. A trip block $k$ has scheduled start time ($\tau^0_k\in \mathbb{R}$) and end time ($\tau^1_k\in \mathbb{R}$) as shown by the blue lines in Fig.~\ref{fig:cabot_schedules_a}\,(b), and a total geographic distance $D_k$. 
It can be seen that trip block schedules are complex and represent operational information that should be included in investment planning. 
We first convert the continuous-time ($\tau$) to discrete-time ($t$). Assume an initial continuous-time reference $\tau_0$, and that there are $S$ representative days and $T_{\text{d}}$ time intervals within each representative day, each with a span of $\Delta T$\footnote{For example, in our numerical study, we consider 8 representative days that cover the four seasons while differentiating weekday and weekend types, using 15-minute resolution intervals, i.e. 96 timesteps per day. For this dataset, $\tau_0$ = 3 AM was chosen.}. We conservatively round down the start times and round up the end times to the nearest interval boundaries, as shown in Fig.~\ref{fig:time_convention}. 
Consider a representative day $s \in [S]$. For each trip block $k \in \mathcal{K}_s$ that belongs to this day, with continuous-time interval $[\tau^0_k, \tau^1_k]$, where $\tau^0_k, \tau^1_k \in \mathbb{R}$, define the discrete-time index $t_k^0, t_k^1 \in \mathbb{N} = \{1,2,...T_{\text{d}}\}$,
\begin{align}
    t^0_k = \left\lfloor \frac{\tau_k^0 - \tau_0}{\Delta T}  \right\rfloor + 1, \quad
    t^1_k = \left\lceil \frac{\tau_k^1 - \tau_0}{\Delta T}  \right\rceil.
    \end{align}
    Furthermore, define parameter matrices, 
\begin{align}
    A_{k}(t) &= \mathbb{1} (t^0_k \leq t \leq t^1_k ),
\\
    U_k(t) &= \mathbb{1} (t = t^0_k ),
\\
    V_k(t) &= \mathbb{1} (t = t^1_k + 1),
\end{align}
where the $A$ matrix describes the intervals when each trip block is active, the $U$ matrix shows when a vehicle must depart the depot for each trip block, and the $V$ matrix shows when a vehicle returns to the depot from a trip block. 

\subsection{Integer-Clustering Formulation of EV Fleets}\label{sec:models_EV}

This section presents the integer-clustering formulation for vehicle fleets. 
The following constraints are enforced to model the fleet operation and dispatch, 
\begin{flalign}
    &\sum_{i\in[I]} b_{ki}=1, \ \ b_{ki} \in \{0,1\}, \quad \forall k, \label{eq_block}
\\
  &n_{is}(t) = N_{i}^\text{v}-\sum_{k \in \mathcal{K}_s}A_{k}(t)b_{ki},\quad \forall i,s,t,\label{eq_depot}
\\
  &\sum_{j\in[J]} m_{ijs}(t)  \leq n_{is}(t),\quad \forall i,s,t, \label{eq_vehicle_depot}
\\
   &\sum_{i \in [I]} m_{ijs}(t)  \leq N_{j}^\text{c},\quad \forall j ,  s,t,\label{eq_vehicle_charging}
\\
    &0\leq p^{\text{v}}_{is}(t)  \leq \sum_{j\in[J]} P_{ij} \cdot m_{ijs}(t),\quad \forall i,s,t,\label{eq_charge_power}
\end{flalign}
where $b_{ki}$ is a binary variable that denotes whether block $k$ is assigned to a type-$i$ vehicle, and \eqref{eq_block} ensures that each block is assigned to exactly one vehicle type. \eqref{eq_depot} enforces that $n_{is}$, the number of type-$i$ vehicles at the depot, equals $N_{i}^\text{v}$, the total number of type-$i$ vehicles, minus those that are en route. The variable $m_{ijs}$ denotes the number of type-$i$ vehicles that are being charged by type-$j$ chargers. 
\eqref{eq_vehicle_depot} requires the number of type-$i$ vehicles that are actively charging at the depot to be upper bounded by the number of type-$i$ vehicles at the depot. \eqref{eq_vehicle_charging} requires that the total number of vehicles that are charging using type-$j$ chargers is upper bounded by $N_{j}^\text{c}$, the number of type-$j$ chargers. \eqref{eq_charge_power} ensures that $p^{\text{v}}_{is}$, the combined charging power for type-$i$ vehicles at the depot is less than or equal to the charging rate times the number of vehicles under charging.

\begin{figure}[h]
    \centering
    \includegraphics[width=0.65\linewidth]{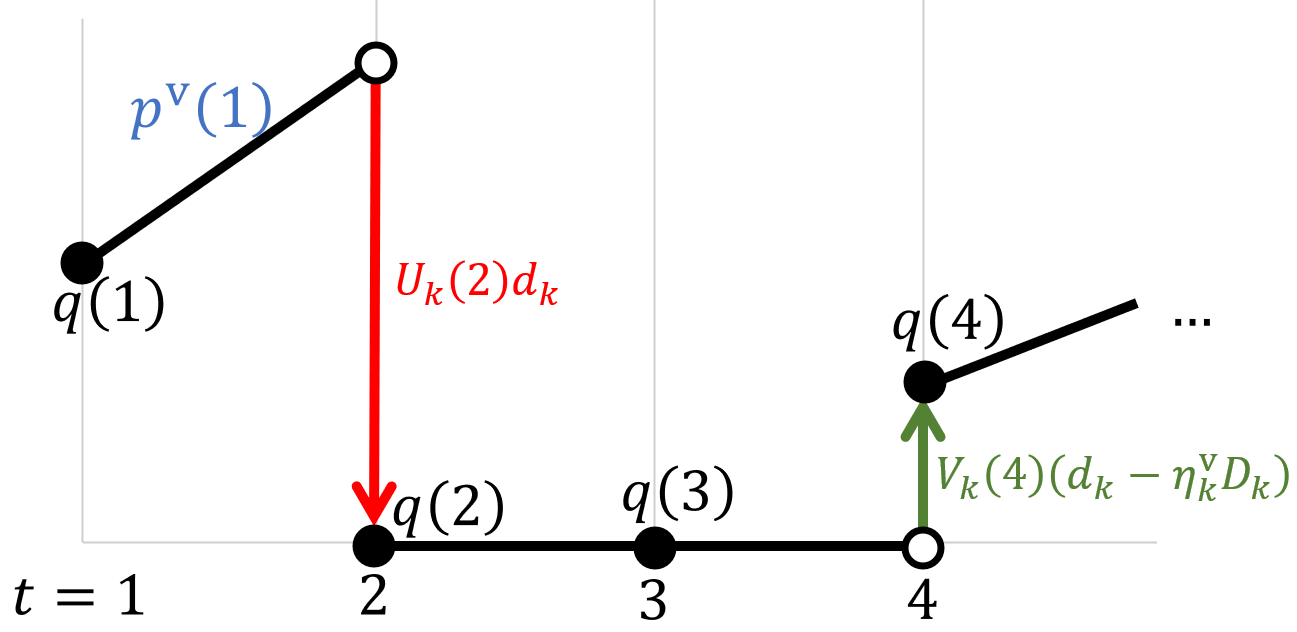}
    \caption{Illustration of dynamics of vehicle SOE, $q(t)$, at the depot. Note the discontinuities when a vehicle leaves or returns to the depot. Here it is assumed that the block is assigned with $b_k=1$, with just one vehicle type.}
    \label{fig:single-vehicle}
\end{figure}

A vehicle's detailed driving behavior, while en route, is encapsulated by the total trip demand $D_k$. 
\eqref{eq_vehicle_soc} tracks (for the $s$-th representative day) the dynamics of $q_{is}(t)$, that is, the combined state-of-energy (SOE) for type-$i$ vehicles that are at the depot,
\begin{align}
        &q_{is}(t+1)  = q_{is}(t)  + p^{\text{v}}_{is}(t)\Delta T  - \sum\nolimits_{k \in \mathcal{K}_s} U_{k}(t+1) d_{ki}  \nonumber
\\
    & \quad + \sum\nolimits_{k \in \mathcal{K}_s} V_{k}(t+1) (d_{ki} - \eta^{\text{v}}_{ki} D_{k} b_{ki}), \quad \forall i,s,t. 
\label{eq_vehicle_soc}
\end{align}
As shown in Fig.~\ref{fig:single-vehicle}, $q_{is}(t)$ varies based on the charging power $p^{\text{v}}_{is}$, the decrease caused by a vehicle leaving the depot, and the increase due to a vehicle returning to the depot carrying unused surplus energy. $\eta^{\text{v}}_{ki}$ is the driving efficiency, and $d_{ki}$ is the energy carried in a type-$i$ vehicle for block $k$.

To account for $d_{ki}$, two formulations can be considered: the first one requires the vehicle being dispatched to carry exactly the amount of energy needed for the trip, 
\begin{flalign}
     &\eta^{\text{v}}_{ki} D_{k} b_{ki}   =d_{ki} \leq R_{i} b_{ki},\ \forall i,k  \quad \text{(more restrictive),} \nonumber
\end{flalign}
where $R_{i}$ is vehicle range. Note that if $b_{ki} = 0$, that is, type-$i$ vehicle is not assigned to block $k$, then $d_{ki}$ is forced to be zero. The alternative formulation allows the vehicle to carry surplus energy than needed by the immediate block,
\begin{flalign}
    &\eta^{\text{v}}_{ki} D_{k} b_{ki} \leq d_{ki} \leq R_{i} b_{ki},\ \forall i,k  \quad \text{(less restrictive).} \label{eq_lower_bound}
\end{flalign}
 
\begin{remark}
 It is clear that the first formulation is more restrictive than in reality where
surplus energy is allowed, while the alternative is less restrictive than in reality
because in this formulation the surplus energy can be redistributed
among the vehicle fleet upon a vehicle returning to the depot.
\end{remark}

Constraint \eqref{eq_vehicle_soc_limit} is a physical limit that ensures the combined energy is upper bounded by the vehicle energy capacity times the number of type-$i$ vehicles at the depot. 

\begin{align}
&0  \leq q_{is}(t)  \leq R_{i} n_{is}(t),\quad \forall i, s, t, \label{eq_vehicle_soc_limit}
\end{align}

For the sake of continuous operation, \eqref{eq_vehicle_soc_ic} requires that the combined energy at the end of the day returns to its initial value,
\begin{align}
     &q_{is}(1) = q_{is}(T_d+1),\quad \forall i,s, \label{eq_vehicle_soc_ic}
\end{align}
where, for notational simplicity, the SOE variables have an additional index $q_{is}(T_d+1)$, and the parameter matrices are augmented as,
$$U_k(T_d+1) = U_k(1),\quad V_k(T_d+1) = V_k(1).$$

Lastly, the non-negative and integer constraints are,
\begin{flalign}
     &N^{\text{v}}_i, N^{\text{c}}_j \in \mathbb{Z}_{\geq0}, \ \forall i, j, s,\\
&m_{ijs}(t) \geq 0,\ \forall i,j,s.  \label{eq_m_positive}
\end{flalign}
Note that, $m_{ijs}$ is modeled as a continuous variable to allow for the flexibility of switching chargers among vehicle types within a time interval.

\begin{remark}
We proved and demonstrated the accuracy and computational efficiency of the integer-clustering formulation in \cite{geng2024integer}. 
A tractable disaggregation can recover an individual solution from the integer-clustering problem.
Mathematical relationships between the integer-clustering, disaggregation, and individual formulations were analyzed. We establishes theoretical lower and upper bounds on the true individual formulation, which underpins a guaranteed performance of the proposed method. Substantial speedups (more than 2000 times for a small instance) with minimal loss in solution quality (less than 0.5\%) were demonstrated. For more details, refer to \cite{geng2024integer}.
\end{remark}
\section{Modeling Of Energy Infrastructures}\label{sec:models}

\begin{figure}[th!]
\centering
\includegraphics[width=8cm]{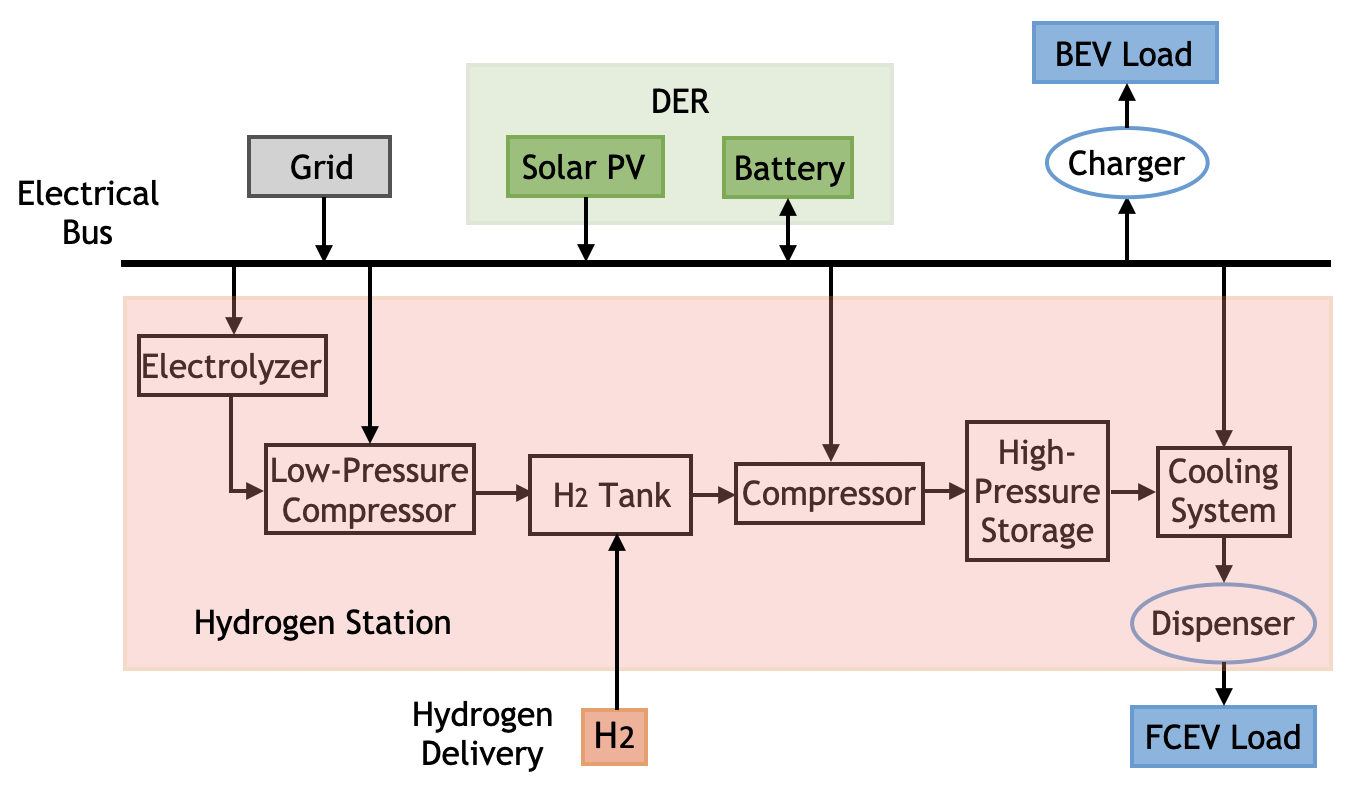}
\vspace{-2mm}
\caption{Schematic of DERs, electrical grid, EV load, charging, and hydrogen fueling infrastructures at the depot.}\label{fig:depot_scheme}
\end{figure}

Figure~\ref{fig:depot_scheme} shows the schematic configuration of a bus depot that supports EV fleets, consisting of DERs, charging facilities, and a hydrogen station.

\subsection{DERs at the Depot}
We consider on-site DERs at the depot such as small-scale renewable generation and battery storage units to reduce the burden on distribution feeders and possibly overall costs \cite{geng2020chance}. The electric power balance equation is given by, 
\begin{flalign}
   &a^\text{pv}_s(t) \overline{p}^{\text{pv}}  + {p}^\text{g}_s(t) = {{p}}^\text{d}_s(t) + {{p}}^{\text{b}+}_s(t) - {{p}}^{\text{b}-}_s(t) + {{p}}^\text{curt}_s(t)\nonumber &&\\
   &\hspace{1cm}  + {{p}}^\text{elz}_s(t)+ {{p}}^\text{lcpr}_s(t) + {{p}}^\text{cpr}_s(t) + {{p}}^\text{cl}_s(t),  \ \ \forall s,t, \label{eq_power_balance_elec}&&
\end{flalign}
where $a^\text{pv}_s(t) \overline{p}^{\text{pv}}$ demotes solar PV generation, ${p}^\text{g}_s(t)$ is grid electricity supply, ${{p}}^\text{d}_s(t)$ is the charging demand of EVs, and ${{p}}^{\text{b}+}_s(t)$, ${{p}}^{\text{b}-}_s(t)$ are the battery charging and discharging power, respectively. A curtailment term, ${{p}}^\text{curt}_s(t)$, is introduced to account for excessive generation from solar PV.  
The terms ${{p}}^\text{elz}_s(t)$, ${{p}}^\text{lcpr}_s(t)$, ${{p}}^\text{cpr}_s(t)$, ${{p}}^\text{cl}_s(t)$ represent the electric power consumed by the local sequence of hydrogen supply:  the electrolyzer (to produce hydrogen), the low-pressure compressor (to compress hydrogen into the hydrogen tank), the compressor (to compress hydrogen from the hydrogen tank to the high-pressure storage buffer), and the cooling system (used when dispensing hydrogen to the vehicle), respectively. 

The energy stored in the battery unit is denoted by ${e}^\text{b}$. Its dynamics are given by,
\begin{flalign}
&{e}^\text{b}_s(t+1) = {e}^\text{b}_s(t) +  \left[{p}^{\text{b}+}_s(t)\eta_c \!-\! {p}^{\text{b}-}_s(t)/\eta_d\right]\! \Delta T,\, \forall s,t, \label{eq_dynamic_bat_e}
\\
&{e}_s^\text{b}(1) = {e}_s^\text{b}(T_d + 1), \ \forall s \label{eq_bat_ic}.
\end{flalign} 
where $\eta_c$ and $\eta_d$ are the charging and discharging efficiencies of the battery, respectively.
\eqref{eq_bat_ic} requires that the energy stored in the battery at the end of each representative day returns to its initial condition.

The stored energy in the battery and the charging/discharging power are subject to physical limits,
\begin{flalign}
&\omega^\text{b}_\text{l} \overline{e}^{\text{b}} \leq {{e}}_s^\text{b}(t) \leq
    \omega^\text{b}_\text{u}  \overline{e}^{\text{b}},  \ \ \forall s,t, \label{eq_cap_bat_e}
\\
    &{{p}}_s^{\text{b}+}(t) + {{p}}_s^{\text{b}-}(t) \leq \overline{p}^{\text{b}}, \ \ \forall s,t, \label{eq_cap_bat_p}
\\
    &0 \leq {{p}}_s^{\text{b}+}(t) \leq M\gamma_s(t),  \ \ \forall s,t,\label{eq_bat_chg}
\\
    &0 \leq {{p}}_s^{\text{b}-}(t) \leq M(1-\gamma_s(t)),  \ \ \forall s,t,\label{eq_bat_dis}
\\
    &\gamma_s(t) \in \{0,1\},  \ \ \forall s,t,\label{eq_bat_gamma}
\end{flalign}
where $\overline{e}_{\text{b}}$ and $\overline{p}_{\text{b}}$ are decision variables representing the energy capacity and power capacity of the battery, respectively. The parameters $\omega^\text{b}_\text{l}$ and $\omega^\text{b}_\text{u}$ in \eqref{eq_cap_bat_e} specify the minimum and maximum energy level of the battery, which is determined by the safe operating range of the battery. $M$ is a large positive constant and the ``big-M'' constraints \eqref{eq_bat_chg}-\eqref{eq_bat_gamma} ensure complementarity of battery charge and discharge, where $\gamma_s(t)=1$ denotes charging.

A typical electricity rate structure for commercial customers consists of energy charge (\$/kWh) and demand charge (\$/kW). The demand charge is based on the peak demand, that is, the highest hourly electricity usage $p_s^{\text{g}}(t)$ over all time intervals during each billing period. The reason for such rate structures is that the cost to supply additional electricity to meet the peak demand is very expensive, and enforcing demand charges incentivizes consumers to decrease their usage during peak hours.
We assume that for the selected representative days, there are in total $L$ groups of season-weekday types with distinct demand charge rates, and $p^\text{pk}_l$ denotes the peak power consumption in the $l$-th group. Let a group $\mathcal{L}_l\subseteq [S]$. We have,
\begin{flalign}
    &p^\text{pk}_l  \geq p_s^{\text{g}}(t)
    ,\quad \forall l\in [L], \ s\in \mathcal{L}_l, \ t\in [T_d].
\end{flalign}

Furthermore, the following engineering and physical constraints are required, 
\begin{flalign}
    &0 \leq {{p}}_s^\text{g}(t) \leq \overline{p}^{\text{g}},  \quad \forall s,t, \label{eq_cap_g_e} \\
    &0 \leq {p}_s^{\text{curt}}(t) \leq
    \overline{p}^{\text{curt}},    \quad \forall s,t, \label{eq_cap_curt}\\
&\overline{p}^{\text{pv}} \geq 0, \quad
\overline{p}^{\text{b}} \geq 0, \quad \overline{e}^{\text{b}} \geq 0. \label{eq_non_neg}
\end{flalign}
We do not consider feeding energy back to the grid for now, leading to $p_s^{\text{g}}(t)\geq 0$ in \eqref{eq_cap_g_e}.
$\overline{p}^{\text{g}}$ and $\overline{p}^{\text{curt}}$ are adjustable parameters that govern the maximum allowable electric power supplied from the distribution system and the maximum allowable curtailment from renewable generation, respectively. Alternatively, $\overline{p}^{\text{g}}$ can be treated as a decision variable, which will be considered in the case study to model the upgrade cost of the distribution system. Non-negativity constraints are given in \eqref{eq_non_neg} for the capacity of solar PV and battery.

\subsection{Hydrogen Fueling Station}
Various technologies can be adopted to meet the hydrogen demand at the fueling station. Specifically, we evaluate two hydrogen supply options: hydrogen imports via truck delivery and on-site generation via electrolysis.

\subsubsection{Hydrogen Delivery by Truck}
The cost of imported hydrogen delivery is represented by an overall price consisting of production and transportation costs.

\subsubsection{On-Site Hydrogen Generation and Storage}
Hydrogen can be generated on-site and stored at the fueling station for local use. Electrolysis is considered in this study, which is an operationally flexible and modular technology that is compatible with different scales of hydrogen demand. The electricity consumption for electrolysis is supplied by the grid or the local renewable generation. In addition, we also allow for flexible electrolytic hydrogen production through the investment in on-site hydrogen storage (low and high pressure) that enables decoupling the operation of the electrolyzer with fueling schedule for the buses at the depot. This operational flexibility is essential for lowering hydrogen production costs by maximizing production during periods of low electricity prices and/or high availability from local DER resources.

\subsubsection{Hydrogen Fueling Process}
As shown in Fig.~\ref{fig:depot_scheme}, 
the compressing and cooling processes consume electricity and lead to associated operational costs.

The mass of stored hydrogen in the (low-pressure) hydrogen tank, ${w}_s^{\text{h}}$, is described by, 
\begin{flalign}
    {w}^{\text{h}}_s(t+1)& ={w}_s^{\text{h}}(t) + \Big(\frac{ {p}_s^\text{lcpr}(t) }{C^{\text{lcpr}}} \! - \! \frac{ {p}_s^\text{cpr}(t) }{C^{\text{cpr}}} \Big) \Delta T  + {w}_s^\text{del}(t), \forall s,t
   ,\label{dynamic_h2}
\end{flalign}
where ${w}_s^\text{del}$ is the mass of the delivered hydrogen, ${p}_s^{\text{lcpr}}$ and  ${p}_s^{\text{cpr}}$ are the electric power consumed by the low- and high-pressure compressors. The parameters
$C^{\text{lcpr}}$ and $C^{\text{cpr}}$ denote the amount of electric energy (kWh) consumed by the low- and high-pressure compressors, respectively, to 
compress 1~kg hydrogen; these factors encapsulate the transformations' efficiency rates. 

The hydrogen produced by the electrolyzer is directly processed by the (low-pressure) compressor in order to be stored in the hydrogen tank \cite{geng2020optimal}. Therefore, the following equality can be established, 
\begin{flalign}
    \frac{{{p}}_s^\text{elz}(t)}{C^{\text{elz}}}  = \frac{ {p}_s^\text{lcpr}(t)}{C^{\text{lcpr}}}, \quad \forall s,t,
\end{flalign}
where the parameter $C^{\text{elz}}$ denotes the amount of electric energy (kWh) consumed by the electrolyzer to produce 1~kg hydrogen.

The mass of stored hydrogen in the (high-pressure) storage buffer, ${w}_s^{\text{bf}}$, is described by, 
\begin{flalign}
    {w}_s^{\text{bf}}(t+1)={w}_s^{\text{bf}}(t) + \Big( 
    \frac{ {p}_s^\text{cpr}(t) }{C^{\text{cpr}}} - 
    \frac{ {p}_s^\text{cl}(t) }{C^{\text{cl}}}\Big) \Delta T,\ \forall s,t,\label{dynamic_h2_buffer} 
\end{flalign}
where ${p}_s^{\text{cl}}$ is the electric power consumed by the cooling system. The cooling system requires $C^{\text{cl}}$~kWh electric energy to process 1~kg hydrogen. 

The following physical and engineering constraints need to be satisfied,
\begin{flalign}
    &\omega^\text{h}_\text{l} \overline{w}^{\text{h}} \leq {w}_s^{\text{h}}(t) \leq
    \overline{w}^{\text{h}}, \quad \forall s,t \label{cap_h2}
\\
    &\omega^\text{bf}_\text{l} \overline{w}^{\text{bf}} \leq {w}_s^{\text{bf}}(t) \leq
    \overline{w}^{\text{bf}},  \quad \forall s,t \label{cap_h2_buff}
\\
    &0 \leq {w}_s^{\text{del}}(t) \leq
    \overline{w}^{\text{del}}, \quad \forall s,t 
    \label{cap_h2_del}
\\
    &0  =   {w}_s^{\text{del}}(t), \quad \forall s, t\notin T_\text{del}, \label{cap_h2_del_time}
\\
    &{w}_s^{\text{h}}(1) ={w}_s^{\text{h}}(T_d+1),  \quad \forall s\label{h2_initial}\\
    &{w}_s^{\text{bf}}(1) ={w}_s^{\text{bf}}(T_d+1), \quad \forall s \label{h2_initial_buff}
\\
    &\overline{w}^{\text{h}} \geq 0, \ \
    \overline{w}^{\text{bf}}\geq 0, 
\end{flalign}
where $\overline{w}^{\text{h}}$ and $\overline{w}^{\text{bf}}$ are decision variables that represent the capacity of the hydrogen tank and buffer. $\omega^\text{h}_\text{l,u}$ and $\omega^\text{bf}_\text{l,u}$ are parameters indicating the minimum and maximum storage levels of the tank and buffer, respectively, to enforce the pressure threshold. ${w}_s^{\text{del}}$ denotes the amount of hydrogen delivered to the station at certain hours in the set $\mathcal{T}_\text{del}$, and it cannot exceed $\overline{w}^{\text{del}}$, the parameter for the maximum deliverable amount of hydrogen at any time. Similar to the battery unit, the storage levels in the hydrogen tank and buffer at the end of the day need to match their beginning values.
$\overline{w}^{\text{h}}$ and $\overline{w}^{\text{bf}}$ are decision variables for the capacity of the hydrogen tank and buffer.

The power capacities of the various components in the hydrogen fueling system, denoted by   $\overline{p}^{\text{elz}}$, $\overline{p}^{\text{lcpr}}$, $\overline{p}^{\text{cpr}}$, and $\overline{p}^{\text{cl}}$, are all decision variables constrained as follows,
\begin{flalign}
 &0 \leq {{p}}_s^\text{elz}(t) \leq \overline{p}^{\text{elz}}, \quad \forall s,t, \label{eq_cap_elz} 
\\
    &0 \leq {p}_s^{\text{lcpr}}(t) \leq
    \overline{p}^{\text{lcpr}}, \quad \forall s,t,
    \label{eq_cap_lcpr}
\\
    &0 \leq {{p}}_s^\text{cpr}(t) \leq \overline{p}^{\text{cpr}}, \quad \forall s,t,\label{eq_cap_cpr} 
\\
    &0 \leq {p}_s^{\text{cl}}(t) \leq
    \overline{p}^{\text{cl}},  \quad \forall s,t,
    \label{eq_cap_cl} 
\\
&\overline{p}^{\text{elz}} \geq 0,   \ \
\overline{p}^{\text{lcpr}} \geq 0, \ \ \overline{p}^{\text{cpr}} \geq 0,\ \ \overline{p}^{\text{cl}} \geq 0. \label{eq_h_posi}
\end{flalign}

\subsection{Carbon Emission Constraint}
We explicitly include a constraint on the overall carbon emissions, i.e. from generating the grid electricity, the production of the delivered hydrogen, and, if any, the combustion of diesel fuel by the benchmark diesel vehicles,  
\begin{flalign}
    \sum_{s\in[S]} \sum_{t\in [T_d]}& N_s^{\text{d}} \left[ 
    \big( p_s^{\text{g}}(t)e^\text{g}_s(t) + {p}_s^\text{df}(t)e^\text{df}\big) \Delta T  + {w}_s^\text{del}(t) e^\text{del}  \right] \leq B, \label{eq_carbon}&&
\end{flalign}
where $N_s^{\text{d}}$ is the number of representative days in a year. $B$ is the parameter for the annual carbon emissions cap. $e^\text{g}_s(t)$, $e^\text{df}$, and $ e^\text{del}$ are emission factors in the unit of kg/kWh for grid electricity, diesel fuel, and delivered hydrogen, respectively. For grid electricity, the emission factor $e_s^\text{g}(t)$ is modeled as a varying parameter with respect to time and representative day, reflecting the marginal generator in the grid.

\subsection{Connecting Energy Model and Vehicle Model}
The energy model and the vehicle model are connected through the electricity demand,
\begin{flalign}
    &{{p}}_s^\text{d}(t) = \sum_{i\in \mathcal{I}_\text{bev}} p^{\text{v}}_{is}(t), \quad \forall s,t. \label{eq_couple_pd}
\end{flalign}

Besides, the total diesel fuel consumption from the benchmark diesel vehicle fleet is given by,
\begin{flalign}
    &{{p}}_s^\text{df}(t) = \sum_{i\in \mathcal{I}_\text{df}} p^{\text{v}}_{is}(t), \quad \forall s,t. \label{eq_couple_ps}
\end{flalign}

The total hydrogen demand ${p}_s^\text{h}$ is related to the vehicle demand and the hydrogen fueling system as follows,
\begin{flalign}
    &{p}_s^\text{h}(t) = \sum_{i\in \mathcal{I}_\text{fcev}} p^{\text{v}}_{is}(t),  \quad\forall s,t, \label{eq_couple_phv}\\
   & {p}_s^\text{h}(t) =  \frac{ {p}_s^\text{cl}(t) }{C^{\text{cl}} } E_{\text{h}}, \quad\forall s,t, \label{eq_couple_phh}
\end{flalign}
where $E_h$ is the kWh energy content per kg hydrogen.

\setlength\tabcolsep{0pt}
\begin{table*}[t] 
\centering
\caption{Economic and technical parameters of battery EV, hydrogen FCEV, and diesel benchmark vehicles \cite{hgacbuy}.}\label{table:EV_parameter}
\begin{tabular}{c || c c c c c c }
\Xhline{2\arrayrulewidth}\\[-0.9em]
\small{\ \ Vehicle Type}  \ & \small{\ Energy capacity ($R_{i}$)\ \ } & \small{\ Range \ }\  & \small{\ \ Capital cost ($c^{\text{v}}_i$)\ } & \small{\ Maintenance cost ($c^{\text{m}}_i$)\ \ } & \small{\ Full charge time\ \ } & \small{\ Life time\ } \\[-0.9em]\\
\Xhline{2\arrayrulewidth}\\[-0.6em]
\small{Proterra ZX5 BEV (short-range)}\  & 225 kWh & 106 miles & \$800,000 & \$0.64/km & 0.45 - 4.5 hours & 12 years  \\[-1em]\\ \hline\\[-0.6em]
\small{Proterra ZX5+ BEV (long-range)} \ & 450 kWh & 197 miles  & \$821,944 & \$0.64/km & 0.9 - 9 hours  & 12 years  \\[-1em]\\\hline\\[-0.6em]
\small{\ New Flyer Xcelsior FCEV\ } & 700 kWh & 350 miles  & \$949,105 & \$0.64/km & 6 - 10 mins  & 12 years \\[-1em]\\\hline\\[-0.6em]
\small{\ New Flyer Xcelsior XDE40 \ } & 5013 kWh & 730 miles  & \$0 & \$0.88/km & 4 - 5 mins  & - \\[-1em]\\
\Xhline{2\arrayrulewidth}
\end{tabular}
\end{table*}

\setlength\tabcolsep{0pt}
\begin{table*}[t] 
\centering
\caption{Economic and technical parameters of DC fast chargers and hydrogen fueling dispenser \cite{johnson2020financial, kushwah2021techno}.}\label{table:charger_parameter}
\begin{tabular}{c || c c c c }
\Xhline{2\arrayrulewidth}\\[-0.9em]
\small{\ \ Charger Type}  \ & \small{\ Power rating ($P_{ij}$)\ \ } & \small{\ Capital cost ($c^{\text{c}}_j$)} \  & \small{\ Life time\ } \\[-0.9em]\\
\Xhline{2\arrayrulewidth}\\[-0.6em]
\small{\ Level-3 DC fast charger \ } & 50 kW & \$37,000 (+ \$22,626 installation cost) & 28 years  \\[-1em]\\ \hline\\[-0.6em]
\small{\ Level-4 DC fast charger\ } & 150 kW & \$45,000 (+ \$22,626 installation cost) & 28 years  \\[-1em]\\ \hline\\[-0.6em]
\small{\ Level-5 DC fast charger\ } & 500 kW & \ \ \$349,000 (+ \$250,000 installation cost)  \ \ & 28 years  \\[-1em]\\ \hline\\[-0.6em]
\small{\ Hydrogen fueling dispenser\ } & 7,000 kW & \$65,000 & 28 years  \\[-1em]\\\hline\\[-0.6em]
\small{\ Diesel fueling dispenser\ } & 72,000 kW & \$0 & - \\[-1em]\\
\Xhline{2\arrayrulewidth}
\end{tabular}
\end{table*}

\section{Case study}\label{sec:results}
We apply the optimization model to a case study of Boston’s public transit bus network and energy facilities at one selected depot, Cabot. All instances were solved using JuMP and Gurobi solver (MIPgap$=0.005$), and computed on a MacBook Pro with 2.3 GHz Intel Core i9 CPU with 8 physical cores and up to 16 logical processors. The MILP problem has 43,244 continuous variables and 7,851 integer variables (among which 7,844 are binary). The various cases were solved in time spans ranging from a few seconds to half an hour.

\subsection{Datasets}\label{sec:data}

\subsubsection{Transportation Data}\label{subsec:fleet_data}
The Massachusetts Bay Transportation Authority (MBTA) publishes public transit schedules on its website \cite{mbta-gtfs}. 
We work with the real dataset of the Cabot bus depot located in south Boston as shown in Figure~\ref{fig:cabot_schedules_a}.
There are in total 1769 trip blocks for the 8 representative days.
The route distances are calculated from the shapefiles provided by MBTA. 
On the weekends, there exist blocks with longer distances. In contrast, there are more trip blocks scheduled on weekdays than on weekends.

\subsubsection{Renewable Generation Data}
Hourly-resolution solar capacity factors were generated using NREL's System Advisor Model \cite{freeman2018system} evaluated for a typical meteorological year at the Cabot Depot coordinates of (42.34, -71.06), and averaged to the 8 seasonal profiles without distinguishing weekday vs. weekend. The on-site solar capacity $\bar{p}^{\text{s}}$ is upper bounded by 1,500 kW, which was estimated by the Cabot Depot's rooftop space measured via Google Maps multiplied by the solar PV footprint of about 100 square feet per kW capacity.

\subsection{Economic and Technical Parameters}\label{sec:economic}

\subsubsection{Parameters for Vehicles and Chargers}\label{subsec:vehicle_para} 
To compare different EV technologies, we consider two BEVs from the Proterra ZX5 product line and a hydrogen FCEV from the New Flyer Xcelsior product line. Their economic and technical parameters are summarized in Table~\ref{table:EV_parameter}. Specifically, the two BEVs feature different ranges, and we refer to them as short-range EV and long-range EV in what follows. In particular, the range of the FCEV is much longer than that of both BEVs, and its full charge time is considerably shorter\footnote{The full charge times in Table~\ref{table:EV_parameter} are estimates. The actual time depends on the power ratings of the charger/dispenser and the energy capacity of EVs.}. 
As a benchmark, we also consider a hybrid diesel-electric bus, the New Flyer Xcelsior XDE40, to model the existing fleets owned by the public transit company. It has a long driving range (730 miles) and energy capacity (5013~kWh). The full charge time is 4-5 mins (equivalent to 72000 kW charging rate). The buses are assumed to be already purchased, so no capital investment is needed. The maintenance cost is \$0.88/km, which is higher than EVs. 
Three types of DC fast chargers with different charging levels are evaluated in the study (Table~\ref{table:charger_parameter}).

\subsubsection{Temperature Effects on Vehicle Driving Efficiency}\label{subsec:temperature} 
\begin{figure}[ht!]
    \centering
    \includegraphics[width=0.97\linewidth]{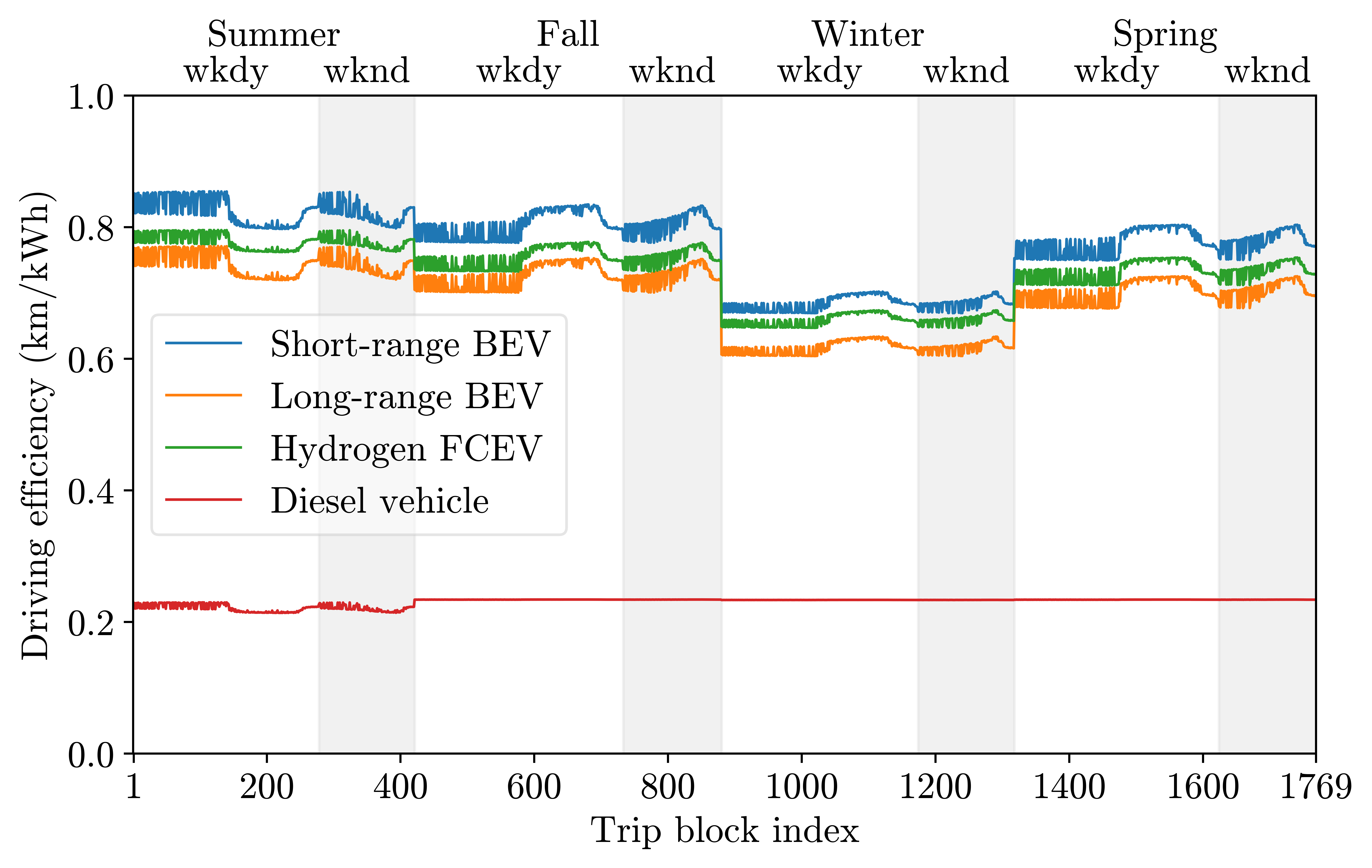}
    \caption{Temperature effects on vehicle driving efficiency $\eta^{\text{v}}_{ki}$. Note the small effect of colder temperatures on diesel driving efficiency, due to the assumed small cold temperature coefficient.}
    \label{fig:temperature}
    \vspace{-0.05cm}
\end{figure}

Temperature can have a significant effect on vehicle driving efficiency which will determine the actual driving range \cite{vculik2021interior}\cite{chiriac2021electric}. To represent this effect, we computed seasonal-diurnal median ambient temperature profiles at 15-minute resolution for the 8 representative days. Then, the temperature differences under or above 65\textdegree F of each interval 
were multiplied by the linear impact coefficients on efficiency according to a systematic study of BEV, FCEV, and diesel buses \cite{henning2020update}\footnote{The hot temperature sensitivity coefficients are (in units of \% impact per $^\circ$F): 0.69 (both BEV types), 0.42 (FCEV), and 0.72 (diesel). The cold temperature coefficients are 0.85 (BEV), 0.69 (FCEV), and 0.01 (diesel).}. Note that cold temperatures tend to have a relatively more detrimental effect on BEV and FCEV efficiencies, whereas high temperatures degrade diesel's efficiency. Each time interval's efficiency multiplier is combined with the vehicles' nominal efficiencies, and integrated over time to calculate the overall driving efficiency of every trip block. As shown in Fig.~\ref{fig:temperature}, the driving efficiency of diesel buses is much lower than that of BEV and FCEV. The short-range BEV has higher driving efficiency than the long-range BEV, partly because the long-range BEV has a heavier battery.

\subsubsection{Electricity Rate Structure}\label{subsec:parameters} 
\begin{figure}[ht!]
\centering
\includegraphics[width=0.97\linewidth]{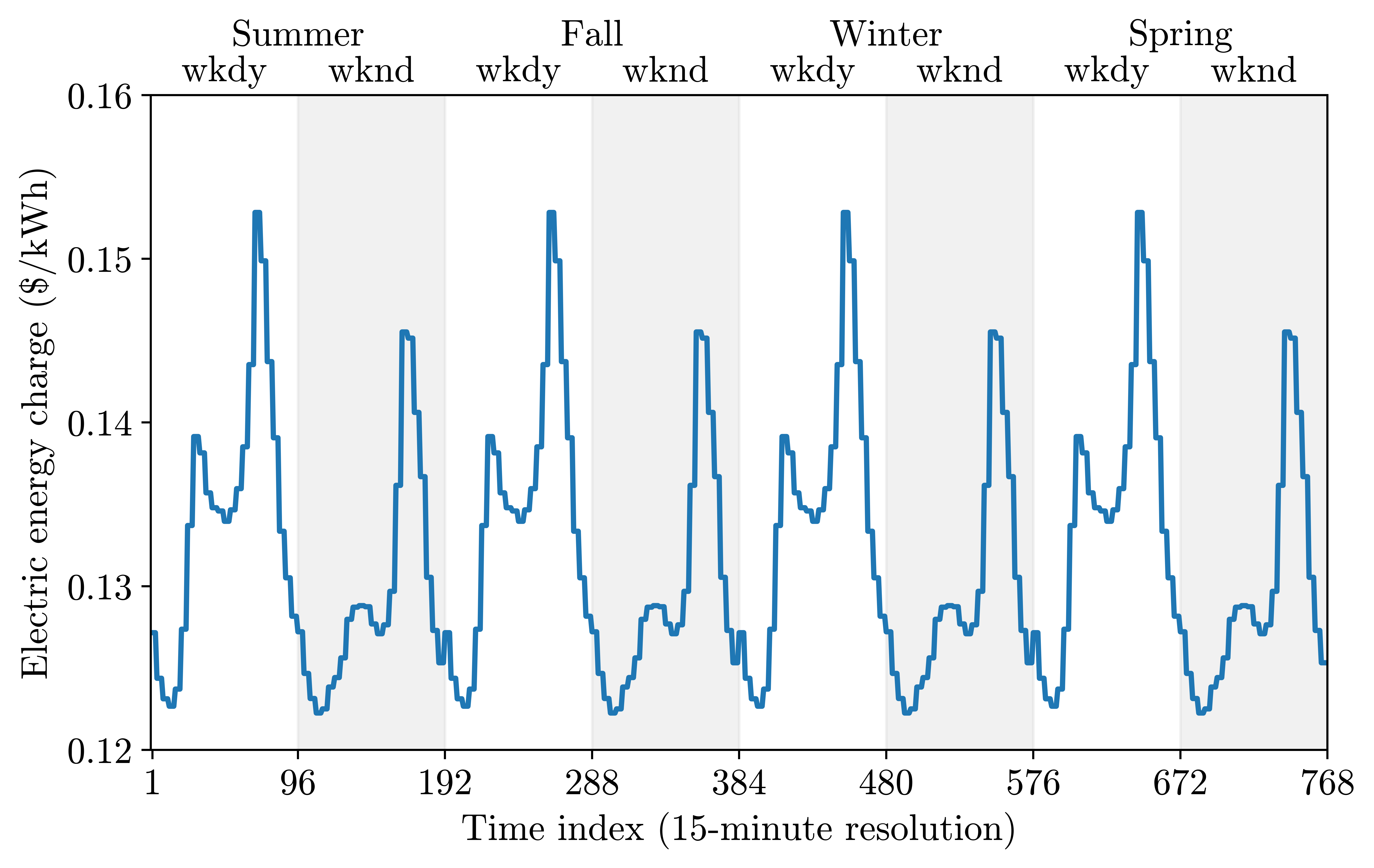}
      \caption{{Synthetic TOU electricity rate ($c_s^\text{g}(t)$) for 8 representative days characterizing various season/weekday types, constructed using ISO-NE wholesale energy prices and scaled to average retail price levels.} }
      \label{fig:isone_tariff}
       \vspace{-0.05cm}
\end{figure}

We adopt a TOU electricity rate based on the ISO New England's wholesale market energy price. 
For each season, we average the energy price over weekdays versus the weekend, and scale and shift the curve to match the average retail electricity price of 13.2~cents/kWh.
As shown in Fig.~\ref{fig:isone_tariff},
the price is higher towards the evening hours of each day when electricity usage increases and is higher on weekdays than weekends. 
Furthermore, a monthly demand charge (in \$/kW) structure is enforced, where 
a higher demand charge (24.09 \$/kW) is enforced in the summer months (Jun-Sep) to incentivize customers to reduce energy consumption during peak demand, while the rest of the months (Oct-May) have a lower demand charge (17.92 \$/kW).

\subsubsection{Parameters for Energy Infrastructures and Hydrogen Fuel}\label{subsec:infra_para} 
The cost parameters for DERs and hydrogen components are summarized in
{\color{black}{Table~\ref{table:infra_parameter}}}. The rest of the parameters for energy conversion are given in {\color{black}{Table~\ref{table:all_tech_parameter}}}. The minimal hydrogen limits are obtained by dividing the minimal ``empty'' pressure of 20 bar by the tank pressures of 350 (low pressure) and 700 
(high pressure) \cite{hua2010compressed}.

\begin{table}[t] 
\centering
\caption{Economic parameters of energy infrastructure.}\label{table:infra_parameter}
\begin{tabular}{c c c c}
\Xhline{2\arrayrulewidth}\\[-0.9em]
Parameter  & Symbol & Value & Unit\\[-0.9em]\\
\Xhline{2\arrayrulewidth}\\[-0.9em]
    Peak grid demand charge (summer) & $c^{\text{pk}}_{1}$ & $24.09$ \cite{openei} \ \ & \$/kW \\[-1em]\\
\hline\\[-1em]
    Peak grid demand charge (non-summer) & $c^{\text{pk}}_{2}$ & $17.92$ \cite{openei} \ \ & \$/kW \\[-1em]\\
\hline\\[-1em]
    Energy cost for grid electricity & $c^\text{g}(t)$ & \ Fig.~\ref{fig:isone_tariff} \cite{iso-ne}  \ \ & \$/kWh \\[-1em]\\
\hline\\[-1em]
    Delivered hydrogen fuel cost (nominal)  & $c^\text{del}$ & $8.0$  \cite{fuelcellbuses2020} \ \ & \$/kg \\[-1em]\\
\hline\\[-1em]
    Diesel fuel cost & $c^\text{df}$ & $4.0$ \cite{eia}  \ \ & \$/gal \\[-1em]\\
\hline\\[-1em]
    Annual maintenance cost for EV fleets & $c^{\text{m}}_\text{EV}$ & $0.64$ \cite{afdc}  \ \ & \$/km \\[-1em]\\
\hline\\[-1em]
    Annual maintenance cost for diesel fleet & $c^{\text{m}}_\text{df}$ & $0.88$ \cite{afdc} \ \  & \$/km \\[-1em]\\
\hline\\[-1em]
    Solar PV annualized capital cost & $c^{\text{s}}$ & $152$ \cite{goodrich2012residential}  \ \ & \$/kW \\[-1em]\\
\hline\\[-1em]
    Battery annualized capital cost  & $c^{\text{b}}$ & $27.4$ \cite{guo2014multi} \ \  & \$/kWh \\[-1em]\\
\hline\\[-1em]
Low-pressure storage annualized capital cost & $c^{\text{h}}$ & $20.9$ \cite{mayer2019techno} \ \  & \$/kg \\[-1em]\\
\hline\\[-1em]
    High-pressure buffer annualized capital cost & $c^{\text{bf}}$ & $33.3$ \cite{rivard2019hydrogen}\ \   & \$/kg \\[-1em]\\
\hline\\[-1em]
Electrolyzer annualized capital cost  & $c^{\text{elz}}$ & $80$ \cite{khan2005pre} \ \  & \$/kW \\[-1em]\\
\hline\\[-1em]
    Low-pressure compressor annualized  & $c^{\text{lcpr}}$ & $15.4$ \cite{mayer2019techno}  \ \ & \$/(kg/h) \\[-1em]\\
    capital cost per H$_2$ throughput capacity & & &\\
\hline\\[-1em]
    High-pressure compressor annualized & $c^{\text{cpr}}$ & $308.3$ \cite{mayer2019techno} \ \ & \$/(kg/h) \\[-1em]\\
     capital cost per H$_2$ throughput capacity & & &\\
\hline\\[-1em]
    Cooling system annualized capital cost & $c^{\text{cl}}$ & $94.1$ \cite{mayer2019techno} \ \ & \$/(kg/h) \\[-1em]\\
      per H$_2$ throughput capacity & & &\\
\Xhline{2\arrayrulewidth}
\end{tabular}
\end{table}

\begin{table}[t] 
\centering
\caption{Technical parameters of energy infrastructure.}\label{table:all_tech_parameter}
\begin{tabular}{c c c c}
\Xhline{2\arrayrulewidth}\\[-0.9em]
Parameter  & Symbol & Value & Unit\\[-0.9em]\\
\Xhline{2\arrayrulewidth}\\[-0.9em]
    Electrolyzer electric efficiency & $C^{\text{elz}}$ & \ \ $41.97$ \cite{maanavi2019energy} \ \ & kWh/kg \\[-1em]\\
\hline\\[-1em]
    Low-pressure compressor electric efficiency & $C^{\text{lcpr}}$ & $0.15$ \cite{doe-hydrogen} & kWh/kg \\[-1em]\\
\hline\\[-1em]
    High-pressure compressor electric efficiency & $C^{\text{cpr}}$ & 3.0 \cite{doe-hydrogen} & kWh/kg \\[-1em]\\
\hline\\[-1em]
    Cooling system electric efficiency & $C^{\text{cl}}$ & 0.2 \cite{doe-hydrogen}  & kWh/kg \\[-1em]\\
\hline\\[-1em]
   Battery charging/discharging efficiency & $\eta_c,\eta_d$  & 0.9,\,0.9\\[-1em]\\
\hline\\[-1em]
   Battery energy lower/upper limit& $\omega^\text{b}_\text{l}$,$\omega^\text{b}_\text{u}$  & 0.2,\,0.9\\[-1em]\\
   \hline\\[-1em]
   Hydrogen storage buffer lower limit&    $\omega^\text{bf}_\text{l}$ & 0.029 \cite{hua2010compressed}\\[-1em]\\
\hline\\[-1em]
   Hydrogen tank lower limit&    $\omega^\text{h}_\text{l}$ & 0.057 \cite{hua2010compressed}\\[-1em]\\
\Xhline{2\arrayrulewidth}
\end{tabular}
\end{table}

\subsection{Benchmark Case: Diesel Bus Fleet}
The Cabot depot currently possesses 55 compressed natural gas buses (New Flyer XN40) and 150 hybrid diesel-electric buses (New Flyer XDE40) \cite{mbta-vehicle-inventory}.  
In this study, we use the New Flyer XDE40 hybrid diesel-electric bus as a benchmark technology and its parameters were given in Section~\ref{sec:economic}.

Applying the optimization model to this case, the optimal operational cost (solved in 4.04 seconds with 0 optimality gap) is \$2.94 million (\$2.826 million diesel fuel cost and \$0.118 million maintenance cost). No investment cost is needed. Annual carbon emission amounts to 7199 (metric) tons of CO$_2$ due to the combustion of diesel fuel.

\subsection{Nominal Case: Fully Electrified Fleet of BEV and FCEV}
We evaluate the two BEVs, one hydrogen FCEV, three DC fast chargers, and local energy infrastructures as potential investment options, as described in Section~\ref{sec:economic}. The optimization model is used to identify the optimal investment strategy for the bus fleet and energy infrastructure. 

\setlength\tabcolsep{0pt}
\begin{table}[ht] 
\centering
\caption{The optimal planning results of the nominal case.}\label{table:results_nominal}
\begin{tabular}{c  c  c  c}
\Xhline{2\arrayrulewidth}\\[-0.9em]
\small{\ \ Vehicle type }  \ & \small{Number} \ \ \ \ & \small{\ \ Charger type\ \ } & \small{Number} \\[-0.9em]\\
\Xhline{2\arrayrulewidth}\\[-0.6em]
\small{\ Short-range EV \ } &\ \ \ \  117 \ \ \ \ \ \ \ \ & \small{\ Level-3,\,5 DC fast charger } & 0   \\[-1em]\\ \hline\\[-0.6em]
\small{\ Long-range EV\ } & \ \ \ \ 31  \ \ \ \ \ \ \ \  & \small{\ Level-4 DC fast charger \ } & 10 \\[-1em]\\ \hline\\[-0.6em]
\small{\ Hydrogen FCEV \ } & \ \ \ \ 0  \ \ \ \ \ \ \ \ & \small{\ Hydrogen dispenser \ } & 0 \\[-1em]\\ 
\Xhline{2\arrayrulewidth}
\end{tabular}
\end{table}

Table~\ref{table:results_nominal} gives the optimal planning results (solved in 10.28 seconds with 0.0013\% optimality gap; The optimal value is \$11.14 million). 
In the nominal case, the fuel cost of delivered hydrogen is assumed to be \$8/kg. At this price, hydrogen FCEVs are not competitive compared to BEVs and they do not appear in the optimal investment plan. The optimization model decides to invest in 117 short-range EVs and 31 long-range EVs, and 10 level-4 DC fast chargers. The (annualized) capital costs of EVs and chargers are \$9.92 million and \$24,152, respectively. Battery is not selected but solar PV is chosen to be built up to the practical limit of 1500~kW (\$228k). The utilization rate of the solar PV turns out to be very high, with negligible PV curtailments. This result also suggests that if the physical space permits more installation of solar panels, a more economical solution may be achieved. 
The cost of electricity usage is \$878.84k, and maintenance of the vehicle costs \$85,801; the sum of which is much less than the operational cost in the benchmark case (2.94 million). This is because the maintenance cost of diesel buses is higher than that of EVs (Table~\ref{table:EV_parameter}). The optimization model also provides detailed operation results. An annual emission of 2213 tons CO$_2$ (from grid electricity) is produced compared to the 7199 tons CO$_2$ in the benchmark case, which is a $69\%$ reduction\footnote{Notably, diesel fuel has a lower emission factor per energy content than the current grid electricity (details provided in Section.~\ref{sec:carbon}). However, since BEVs have much higher efficiency than the hybrid diesel bus (Tabel~\ref{table:EV_parameter}), converting the current fleet from hybrid diesel to BEV achieves a significant carbon reduction.}. 

\begin{figure}[h]
\begin{center}

    \includegraphics[width=0.97\linewidth]{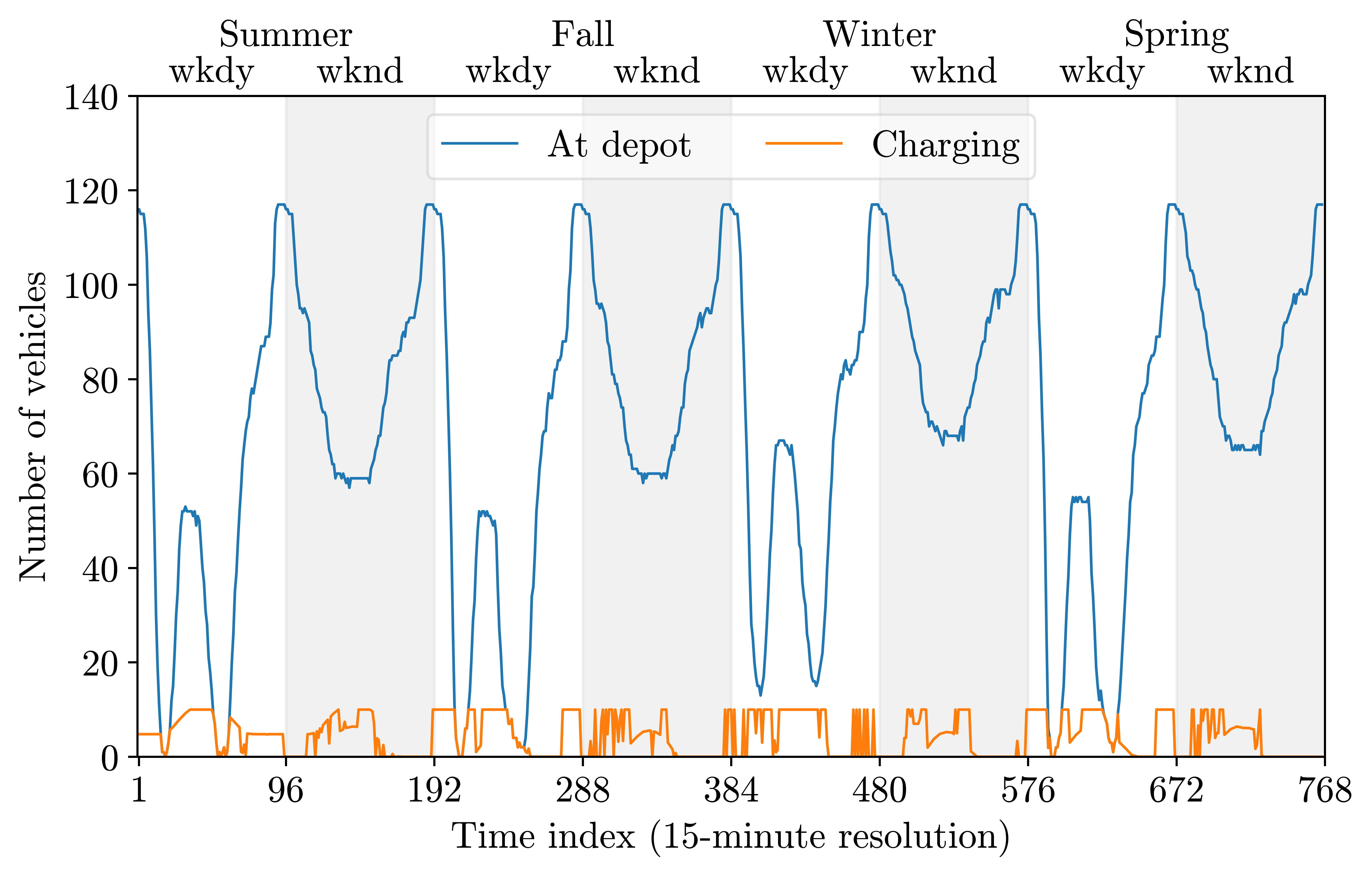}

\end{center}
      \caption{{Number of short-range EVs at the depot and those that are being charged for the nominal case. 
      }} 
\label{fig:operation_nominal_short}
       \vspace{-0.05cm}
\end{figure}

\begin{figure}[h]
\begin{center}
    \includegraphics[width=0.97\linewidth]{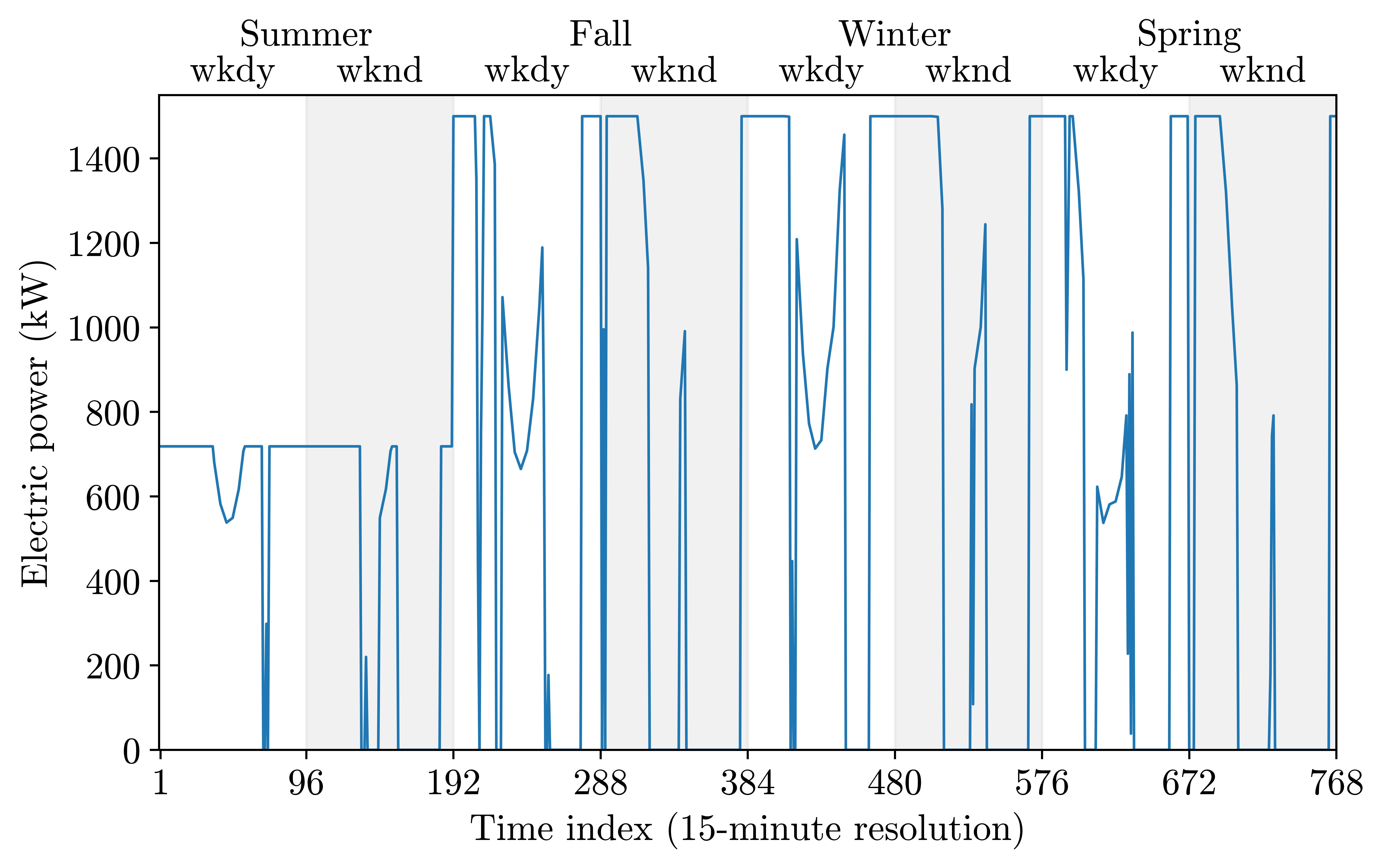}

\end{center}
      \caption{{Total charging power from the grid. Note the lower summer peak usage due to a higher summer peak demand charge. } }
\label{fig:operation_nominal_grid}
       \vspace{-0.05cm}
\end{figure}

Figs.~\ref{fig:operation_nominal_short} 
shows the number of vehicles at the depot versus those that are being charged for the short-range EVs.
We can observe that, most of the time, only a small portion of the vehicles at the depot are being charged. Smart charging requires fewer chargers to be built. 
The total charging power from the grid is shown in Fig.~\ref{fig:operation_nominal_grid}. Due to the higher demand charge in the summer months, the peak demand in summer is lower than in the other seasons. Also, since the TOU energy price is higher toward 8 pm at night (Fig.~\ref{fig:isone_tariff}), the model chooses not to draw electricity from the grid around that hour but rather charges later in the evening.

\subsection{Sensitivity Analysis on Hydrogen Fuel Cost}
We perform a sensitivity analysis on the cost of delivered hydrogen. It turns out that hydrogen FCEV can only compete with BEVs when the price drops to about \$1/kg
, the planning results of which are given in Table~\ref{table:results_hydrogen} 
(solved in 58.96 seconds with 0.0415\% optimality gap; The optimal value is \$11.12 million). This indicates that substantial cost reductions are needed for hydrogen to be a viable alternative to BEVs. For context, the Inflation Reduction Act in the U.S. provides a production tax credit of up to \$3/kg for hydrogen production facilities that meet certain requirements. DOE's Hydrogen Shot initiative, launched in 2021, seeks to reduce the production cost of clean hydrogen by 80\% to \$1/kg in one decade.

\setlength\tabcolsep{0pt}
\begin{table}[h] 
\centering
\caption{The optimal planning results when the hydrogen fuel cost decreases to \$1/kg.}\label{table:results_hydrogen}
\begin{tabular}{c  c  c  c}
\Xhline{2\arrayrulewidth}\\[-0.9em]
\small{\ \ Vehicle type }  \ & \small{Number} \ \ \ \ & \small{\ \ Charger type\ \ } & \small{Number} \\[-0.9em]\\
\Xhline{2\arrayrulewidth}\\[-0.6em]
\small{\ Short-range EV \ } &\ \ \ \  117 \ \ \ \ \ \ \ \ & \small{\ Level-3,\,5 DC fast charger } & 0   \\[-1em]\\ \hline\\[-0.6em]
\small{\ Long-range EV\ } & \ \ \ \ 12  \ \ \ \ \ \ \ \  & \small{\ Level-4 DC fast charger \ } & 7 \\[-1em]\\ \hline\\[-0.6em]
\small{\ Hydrogen FCEV \ } & \ \ \ \ 19  \ \ \ \ \ \ \ \ & \small{\ Hydrogen dispenser \ } & 1 \\[-1em]\\ 
\Xhline{2\arrayrulewidth}
\end{tabular}
\end{table}

At the \$1/kg hydrogen fuel price, the design has a
vehicle cap cost of \$11.12 million (117 short-range EVs, 12 long-range EVs, and 19 hydrogen FCEVs) and a charger cap cost of \$19,228 (7 level-4 DC fast charger and 1 hydrogen fueling dispenser).
Electricity cost is \$581.25k, and delivered hydrogen fuel cost is \$73,160.
Maintenance cost is \$85,801. The battery is not selected, and solar PV is built up to the capacity of 1500 kW. The local hydrogen generation method through electrolysis is not chosen ($\overline{p}^{\text{elz}}=$0, $\overline{p}^{\text{lcpr}}=$0) because it is more costly than hydrogen delivery.
Hydrogen storage and fueling infrastructures are built at the depot, which include a hydrogen tank ($\overline{w}^{\text{h}}=$114.94 kg), compressor ($\overline{p}^{\text{cpr}}=$39.02 kW), high-pressure storage buffer ($\overline{w}^{\text{bf}}=$60.57 kg), and cooling system ($\overline{p}^{\text{cl}}=$5.08 kW). 

\begin{figure}[h!]
\begin{center}
    \includegraphics[width=0.97\linewidth]{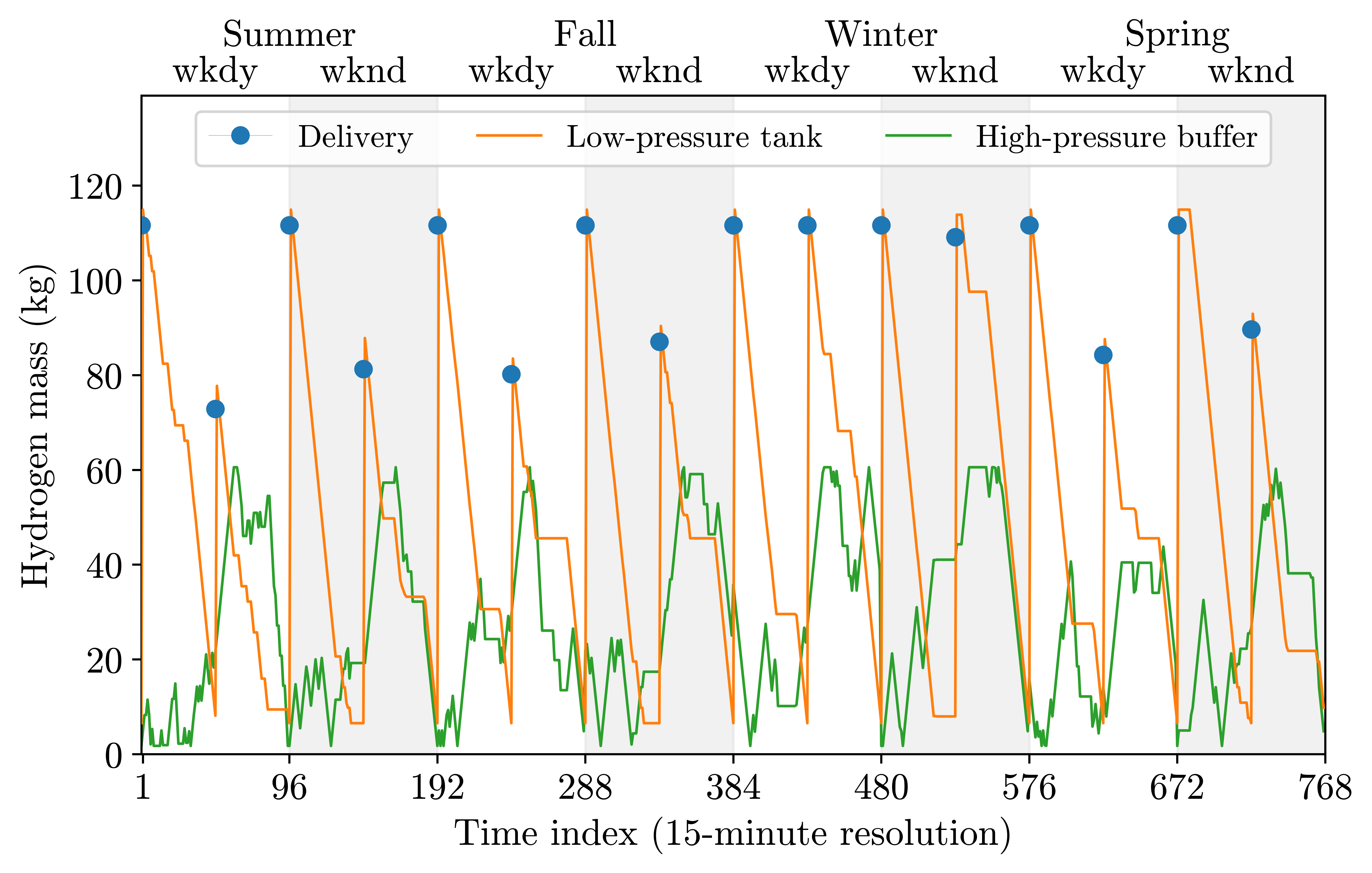}
    
\end{center}
      \caption{{Dynamics of hydrogen mass delivered by truck, stored in the low-pressure tank, and in the high-pressure buffer. Note that the low-pressure tank's mass exhibits spikes in the intervals immediately following hydrogen deliveries.}  }
\label{fig:operation_hydrogen}
       \vspace{-0.05cm}
\end{figure}

Figure~\ref{fig:operation_hydrogen} shows the dynamics of hydrogen mass in the low-pressure tank and high-pressure buffer. The yellow spikes denote the delivered hydrogen which happens twice a day, to be saved in the low-pressure tank for later utilization throughout the day. 

\subsection{Analysis of Carbon Emission Constraint}\label{sec:carbon}
The model tracks total carbon emissions, including those due to the combustion of fuels and the emissions from generating grid electricity. 
We assume that the delivered hydrogen (at \$8/kg) has no carbon emissions and that the diesel combustion factor $e^\text{dies}$ is 0.25 kg CO$_2$/kWh diesel energy. The emission factor of grid electricity $e_s^\text{g}(t)$ is a time-varying parameter based on the levelized long-run emission factors from NREL's Cambium dataset \cite{nrelcambium}. The monthly-hourly values are averaged by seasons and re-sampled into our 768-interval time series (in the order of summer, fall, winter, and spring).

We impose a constraint on the carbon emissions and analyze how it affects the system design. Recall that the diesel bus benchmark gives a 7199 tons CO$_2$ emission annually. The optimization model is re-run with diesel buses included as an option, assuming no capital costs for the diesel buses. Table~\ref{table:carbon_result} shows how the vehicle mix changes when we progressively tighten the carbon constraint. As expected, the number of diesel buses is monotonically decreasing and the total (annualized) cost is increasing.
When the system is fully decarbonized, the fleet is composed of 42 short-range BEVs and 106 hydrogen FCEVs. Note that the BEVs are supported by renewable generation. 

\setlength\tabcolsep{0pt}
\begin{table}[] 
\centering
\caption{Optimal design results under various carbon emission cap.}\label{table:carbon_result}
\begin{tabular}{c || c c c c c c}
\Xhline{2\arrayrulewidth}\\[-0.9em]
\small{\ \ Carbon Cap}  \ & \ \ 100\% \ \  & \ \ 80\% \ \ & \ \ 60\%  \ \ & \ \ 40\%  \ \ & \ \ 20\%  \ \ & \ \ 0\%  \ \ \\
\Xhline{2\arrayrulewidth}\\[-0.6em]
\small{\ Short-range EV\ } & 0 & 9 & 17 & 22 & 22 & 42\\[-1em]\\ \hline\\[-0.6em]
\small{\ Long-range EV\ } & 0 & 6 & 3 & 0 & 0 & 0 \\[-1em]\\ \hline\\[-0.6em]
\small{\ Hydrogen FCEV\ } & 0 & 0 & 12 & 29 & 55 & 106 \\[-1em]\\ \hline\\[-0.6em]
\small{\ Diesel vehicle\ } & 148 & 133 & 116 & 97 & 72 & 0 \\[-1em]\\\hline\\[-0.6em]
\small{\ Total cost (million)\ } & 2.94  & 3.63 & 4.65 & 6.08 & 8.05 & 13.33 \\[-0.6em]\\
\Xhline{2\arrayrulewidth}
\end{tabular}
\end{table}

\subsection{Analysis on the Cost of Distribution System Upgrade}
Finally, we perform a simple analysis on the cost of distribution system upgrades due to the increased burden from EVs. Under the setup in the nominal case, instead of allowing infinite electricity power to be supplied from the grid, we assume an upper bound of 1~MW. If the EV fleets demand more electric power than this amount, an upgrade cost of \$500/kW annually is incurred.
Although more detailed analysis will be carried out in the future for the impacts on distribution system upgrade needs, this preliminary analysis shows that with the given assumptions, the upgrade cost is negligible compared to the investment cost for EV fleets.

\section{Conclusion}\label{sec:conclusion}
This paper makes a novel attempt at representing the complex dynamics between power systems and future electric vehicle (EV) fleets to determine the optimal fleet composition and corresponding energy infrastructure. A computationally efficient integer-clustering method enables the modeling of large-scale EV fleet operations. Applying the optimization model to a real-world case study, we provide realistic insights into the future evolution of interconnected electricity-hydrogen-transportation systems. It shows that converting the current hybrid diesel fleet to EVs achieves a significant carbon reduction. Distributed energy resources such as solar panels can reduce the overall cost and carbon emissions. Results also show that, for hydrogen EVs to be a viable alternative to battery EVs, substantial cost reductions are needed for hydrogen fuels and infrastructure. Furthermore, enforcing carbon emission constraints can have a significant impact on the system design. For future work, we will consider multiple charging depots, en-route charging infrastructure, and energy distribution networks. We plan to extend the framework to consider stochastic transportation demands, and will look into the effects of different electricity and hydrogen pricing structures on the planning and operation results.

\bibliographystyle{IEEEtran}
\bibliography{RefTransp_202304}

\newpage

\vfill

\end{document}